%% file: FinalDraft.tex
\documentclass[useAMS,usenatbib]{mn2e}

\usepackage{longtable}
\usepackage{graphicx}

\newcommand\apj{ApJ}
\newcommand\aj{AJ}

\newcommand\apjs{ApJS}
\newcommand\aap{A\&A}

\newcommand\pasp{PASP}

\newcommand\apss{Ap\&SS}
\newcommand\memsai{MmSAI}

\title[YSOs in the Lupus Molecular Clouds]{Spectroscopic properties of
  Young Stellar Objects in the Lupus Molecular Clouds}
\author[A. Mortier et al.]{Annelies Mortier$^{1,2}$\thanks{E-mail:
    amortier@astro.up.pt}, Isa Oliveira$^{2}$ and Ewine F. van
  Dishoeck$^{2,3}$\\
$^1$ Centro de Astrof\'{\i}sica \& Faculdade de Ci\^encias, Universidade do
  Porto, Rua das Estrelas, 4150-762 Porto, Portugal\\
$^2$ Leiden Observatory, Leiden University, P.O. Box 9513, 2300 RA
  Leiden, The Netherlands\\
$^3$ Max-Planck-Institut f\"ur Extraterrestriche Physik, P.O. Box
  1312, D-85741 Garching, Germany}

\begin{document}

\date{Accepted 2011 August 2.  Received 2011 July 11 ; in original form 2011 April 11}

\pagerange{\pageref{firstpage}--\pageref{lastpage}} \pubyear{2011}

\maketitle

\label{firstpage}

\begin{abstract}
The results of an optical spectroscopic survey of a sample of young
stellar objects (YSOs) and pre-main sequence (PMS) stars in the Lupus
Clouds are presented. 92 objects were observed with VLT/FLAMES. All of
those objects show IR excess as discovered by the \textit{Spitzer}
Legacy Program ``From Molecular Cores to Planet-Forming Disks''
(c2d). After reduction, 54 spectra with good signal-to-noise ratio are
spectrally classified. Effective temperatures and luminosities are
derived for these objects, and used to construct H-R diagrams for the
population. The sample consists mostly of M-type stars, with $10\%$
K-type stars. Individual ages and masses are inferred for the objects
according to theoretical evolutionary models. The mean population age
is found to be between 3.6 and 4.4 Myr, depending on the model, while
the mean mass is found to be $\sim$0.3 $M_\odot$ for either model.
Together with literature data, the distribution of spectral types is
found to be similar to that in Chamaeleon I and IC348. The H$\alpha$
line in emission, found in $49\%$ of the sample, is used to
distinguish between classical and weak-line T Tauri stars. $56\%$ of
the objects show H$\alpha$ in emission and are accreting T Tauri
stars. Mass accretion rates between $10^{-8}$ and $10^{-11}$
M$_{\odot}$yr$^{-1}$ are determined from the full width at $10\%$ of
the H$\alpha$ peak intensity. These mass accretion rates are, 
within a large scatter, consistent with the $\dot{M}_{ac} \propto M^2$ 
relation found in the literature.
\end{abstract}


\section{Introduction}

\textit{Lupus Molecular Clouds} is the generic denomination of a
loosely connected concentration of dark clouds and low-mass pre-main
sequence stars located in the Scorpius-Centaurus OB association at
$16^h20^m < \alpha < 15^h30^m$ and $-43^{\circ} < \delta <
-33^{\circ}$. Due to its large size, close distance ($d$ = 150--200
pc) and substantial mass of molecular gas, the Lupus Clouds have been
subject of many studies at all wavelengths over the years
(e.g. \citealt{Hughes94,Merin08}; see \citealt{Comeron08} for a
review; \citealt{Comeron09,Tothill09}).

Lupus is one of the five clouds selected by the \textit{Spitzer}
Legacy Program \textit{``From Molecular Cores to Planet-Forming
  Disks''}, also referred to as c2d (Cores to Disks, Evans et
al. 2003). Using 400 hours of observations and all three instruments
of \textit{Spitzer}, the c2d program studies the process of star and
planet formation from the earliest stages of molecular cores to the
epoch of planet-forming disks. The five observed clouds cover a range
of cloud types broad enough to study all modes of low-mass star
formation, and large enough to allow statistical conclusions. A rich
population of low-mass young stellar objects (YSOs) still surrounded
by their circumstellar material has been discovered by {\it Spitzer}
in these clouds \citep{Harvey07a,Harvey07b,Merin08,Evans09}.

Disks around YSOs, called protoplanetary disks, consist mainly of dust
and gas. Dust particles of sub-$\mu$m size dominate the disk opacity,
making it easily observable at infrared and (sub-)millimeter
wavelengths by re-emitting some of the received stellar
radiation. Dust emission is also temperature dependent with colder
dust emitting at longer wavelengths than warm dust. Protoplanetary
disks evolve in time, ending up in different scenarios like a
planetary system or a debris disk, that may or may not also harbor
planets.

The stellar radiation spectrum depends on the effective temperature of
the star. Because of the properties of dust emission, the IR excess in
a spectral energy distribution (SED) of this star+disk system provides
information about the geometry and properties of the dusty disk. To
separate star and disk emission, the stellar characteristics need to
be known. The inner regions of the circumstellar disk are disrupted by the
stellar magnetosphere. The stellar magnetic field creates channels
through which material can flow from the disk onto the star. This
produces atomic lines from hydrogen, calcium, oxygen, etc. The
strongest emission line is H$\alpha$ at 6562.8 \AA{}.

The Lupus complex is one of the largest low-mass star forming regions
on the sky, with four main star forming sites, referred to as Lupus
I-II-III-IV. Furthermore, Lupus III contains one of the richest
associations of T Tauri stars (see \citealt{Comeron08} for a
review). The work presented here concentrates on data from Lupus I,
III and IV because those were the regions observed by the c2d
program. Their star formation rates are 4.3, 31 and 4.5 M$_{\odot}$
Myr$^{-1}$, respectively \citep{Evans09}. Distances of $150\pm 20$ pc
for Lupus I and IV and $200\pm 20$ pc for Lupus III are assumed in
this work \citep{Comeron08}.

We present an optical spectroscopic survey designed to characterize
the young stellar population of Lupus I, III and IV, as observed by
the c2d program. A similar study on a subsample is performed by 
\citet{Alcala11}. Their data from an X-shooter survey are still 
preliminary, but suitable for a parallel study to ours. 
Section \ref{sample} describes the selection criteria
for the sample and Section \ref{data} the observations and data
reduction. Using libraries of standards from the literature, spectral
types and effective temperatures are obtained in Section
\ref{SpT}. Combined with additional photometric data, the SEDs of the
objects can be built. In Section \ref{SED}, the stellar and disk
luminosities are calculated from the SEDs. In Section \ref{HR}, the
stars are placed in Hertzsprung-Russell diagrams and individual masses
and ages are derived based on theoretical tracks. The mass accretion
rates, determined through the H$\alpha$ line, are presented in Section
\ref{Halpha}. In Section \ref{disc} the young stellar population of
Lupus is discussed in context with other regions. Finally, in Section
\ref{conclus}, the conclusions from this work are stated.

\section{Sample selection}\label{sample}

\citet{Merin08} used the \textit{Spitzer} c2d point source catalog to
identify the YSO population in the Lupus clouds. Objects are
classified as YSO if they show an IR excess in the SED. To obtain an
optimal separation between young stars, background galaxies and
Galactic post-AGB stars, \citet{Merin08} used the selection criteria
developed by the c2d team on its official point source catalog
\citep{Evans07}. The method relies on an empirical probability
function that depends on the relative position of any given source in
several color-color and color-magnitude diagrams where diffuse
boundaries have been determined.

The list of YSOs from \citet{Evans07} was then adapted by
\citet{Merin08}. Visual inspection was performed to subtract suspected
galaxies or binaries, leaving the list with 94 YSOs.
The final list of \citet{Merin08} was merged with 65 pre-main sequence
(PMS) stars and PMS candidates. Here, the term PMS star is used for
other objects added to the list whose youth had already been confirmed
using other observational techniques, mainly optical spectroscopy.  If
an object has not been spectroscopically confirmed as young but it was
selected by its optical and near-IR photometry as such, it is labeled
as a PMS candidate. This final list of 159 young objects is used for
the observations presented here.

\section{Observations and data reduction}\label{data}

The data presented here were taken in the second half of the nights of
20 -- 25 February 2008 with the Very Large Telescope (VLT) and the
instrument FLAMES/GIRAFFE (ID: 080C.0473-A, PI: Oliveira). The
instrument was used in the MEDUSA mode, with wavelength coverage of
6437 -- 7183 \AA{}, and spectral resolution of 0.79 \AA{}. This
wavelength range was chosen for containing temperature sensitive
features, useful for spectral classification. Additionally, it covers
the H$\alpha$ line, an accretion diagnostic. MEDUSA has 135 fibers
available, each with an aperture of 1.2\arcsec. In total, 250 stars in
19 fields were observed (on average: 14 stars per field). Of those,
158 are field stars, leaving the Lupus science sample with 92
objects. An overview of the observations can be found in Table
\ref{TabID}. The different exposure times are adjusted to the mean
magnitude of the objects in a given field to avoid saturation of the
brightest sources.

The VLT data pipeline, GASGANO, was used for the data reduction. For
each observation night, a set of 5 dark frames, 3 flat-fields and 1
arc frame are produced. GASGANO performs bias subtraction,
flat-fielding and wavelength calibration. Flux calibration 
was not required for the science goals since the bands used for 
classification are close to each other. Further spectral extraction 
was performed within {\rm IDL}. In each observed field, unused fibers
were placed in random ``empty'' sky positions. Those positions are not
really ``empty'' sky, since the cloud itself can contribute to the sky
with a lot of emission lines. In each field, all the sky spectra are
combined within {\rm IDL} into one master sky spectrum, with a
2$\sigma$ clipping. This combined sky is then subtracted from the
science spectra to obtain spectra that are ready for analysis.

The spectra are sorted into four categories: good spectra,
non-detection with a mean flux around zero, featureless spectra and
not useful spectra (marked as G, U, F and O respectively in Table
\ref{TabID}). The two spectra classified as not useful for our science
purpose are objects \# 14 and 53. Object 14 was discovered to be a
galaxy (see section \ref{special}) while object 53 was observed with a
broken fibre, resulting in an unclassifiable spectrum. The
non-detections (marked `U', 23 objects) and featureless (marked `F',
13 objects) spectra could also not be spectrally classified. This
leads to 54 good classifiable spectra, out of which 26 have a
signal-to-noise ratio greater than 20.

\section{Spectral Classification}\label{SpT}

\subsection{Method}

From the obtained spectra, the YSOs were classified by comparing their
spectral features with libraries of standard stars. To match the high
resolution of the FLAMES spectra, high resolution standards from
\citet{Montes98} were used. However, standards for all spectral types
are not available in this library, so that only a range of spectral
types can be derived from it. To further the analysis and narrow down
the spectral type, a low resolution (3.58 \AA{}) library was used as
well (G. Herczeg, private communication, 2010). Once a spectral type
range has been determined from the high resolution standards, the low
resolution library is used for a finer determination due to the
availability of standards of nearly all spectral types in this
library.

The most prominent features in late type stars are the Titanium Oxide
(TiO) absorption lines at 7050 -- 7150 \AA{}. Mostly K-type and M-type
stars have these features because they are cold enough for TiO to
exist. The TiO feature at these wavelengths shows three absorption
bands, which are deeper for colder objects. The Li {\sc i} line in
absorption at 6707 \AA{} is also a common feature in young stars. 
This line was detected in some objects (see Table \ref{TabSpTAV}), 
but was not used to spectrally classify the objects.

\begin{figure}
\begin{center}
\includegraphics[angle=270,width=9cm]{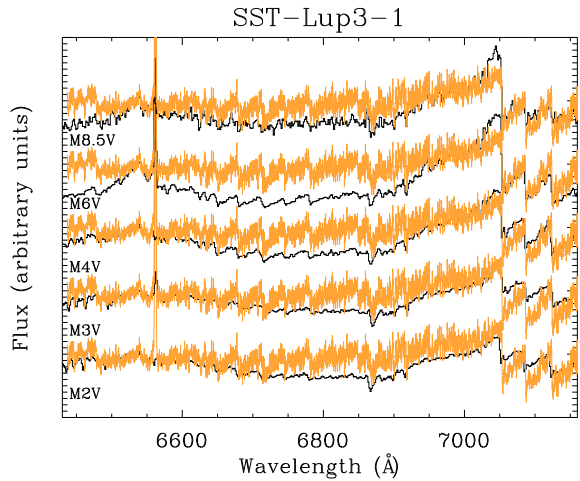}
\caption{Example of the classification method of SST-Lup3-1 with high
  resolution standards. The black curves represent the high resolution
  standard models and the orange curves the science spectrum. At the
  left of each standard its spectral type is indicated. The range of
  plausible spectral types is determined to be M4 - M8.5}
\label{FigHighR}
\end{center}
\end{figure}

\begin{figure}
\begin{center}
\includegraphics[width=9cm]{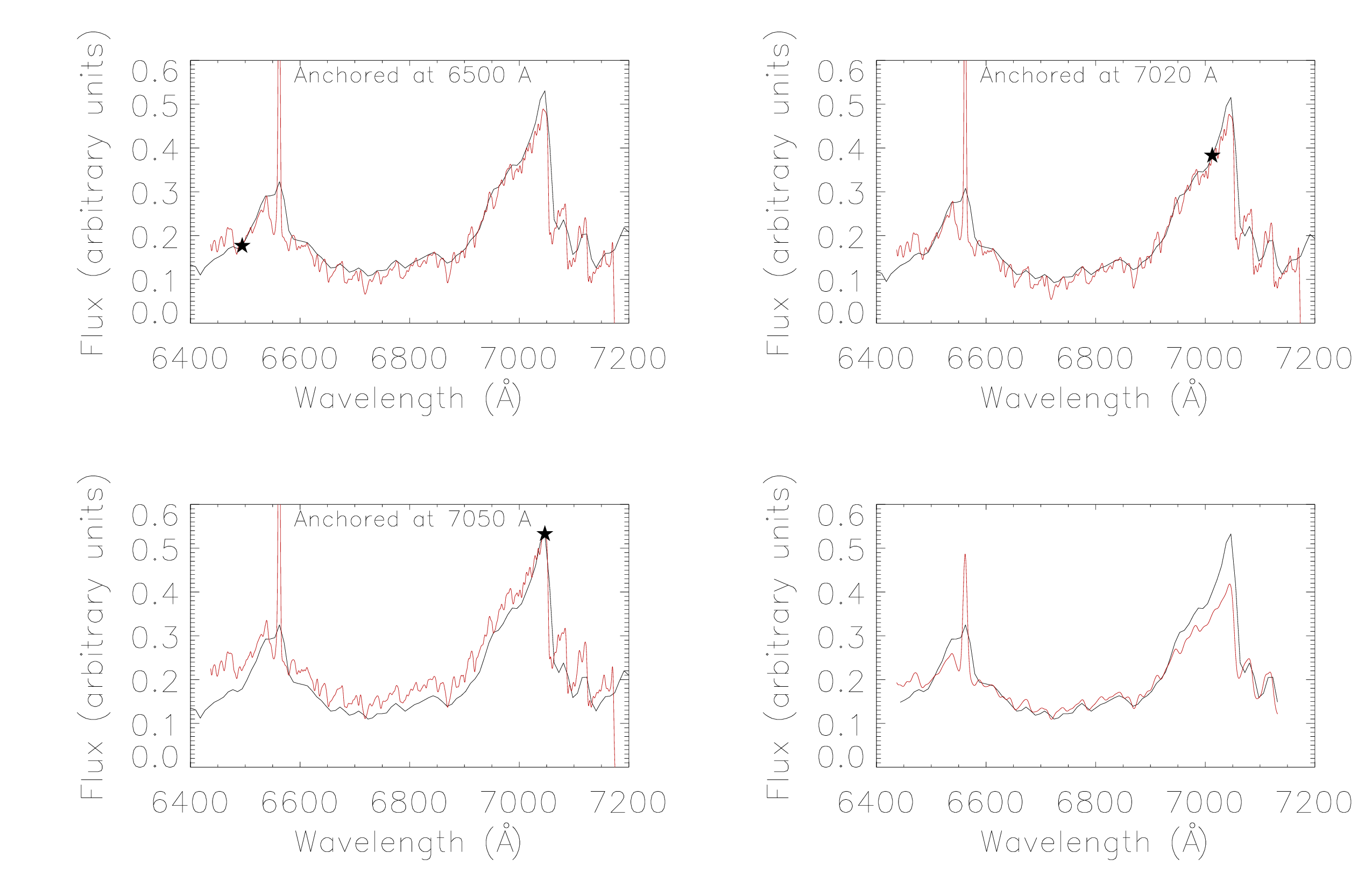}
\caption{Example of the classification methods of SST-Lup3-1 with the
  low resolution standard of a M6 main sequence star. The first three
  panels show the first method where the spectra are anchored at
  different wavelengths. The last panel shows the second method.}
\label{FigLR}
\end{center}
\end{figure}

To compare with the high resolution standards, the science spectra are
normalized (divided by the continuum) and overplotted on the
normalized standard spectra of different spectral types (see Figure
\ref{FigHighR}). A spectral range is visually determined by the best
match for the TiO bands. The method for the low resolution standards
is somewhat different because that library is not normalized. First,
the resolution of the science spectra is lowered to that of the
standard ($3.59$\AA{}). This is done by convolving the spectrum with a
Gaussian profile.  The comparison of the spectra with the low
resolution standards then happens in two ways. One way consists of
scaling the spectra to the standards by anchoring their fluxes at
certain wavelengths (6500, 7020 and 7050 \AA{}). An example is shown
in Figure \ref{FigLR} (top left, top right and bottom left). In the
alternative method a scale factor is determined by first dividing both
spectra and then taking the mean of those values, ignoring extreme
features (like H$\alpha$ emission lines) in the process. This factor
is then used to scale the original spectra to the standards (see
bottom right panel of Figure \ref{FigLR}). Both methods agree very
well, producing four different plots for each spectral type within the
determined range. The correct spectral type is again visually
determined by the best match at the TiO bands. The typical uncertainty
in the spectral classification is one sub-class.

\subsection{Special spectra}\label{special}

Some spectra present special features besides the temperature
sensitive ones used for spectral classification. These objects were
inspected more closely and are differentiated in three special types
of spectra:

\begin{itemize}

\item Ten spectra show [He {\sc i}] emission lines at 6677.6 and
  7064.6 \AA{} (objects \# 7, 52, 76, 82, 83, 84, 85, 87, 91, 93 - see
  examples in Figure \ref{FigHeI}). All of these objects also show
  strong H$\alpha$ in emission. Strongly accreting objects produce
  emission lines other than H$\alpha$, such as this line. It is
  concluded that these lines are a sign of accretion, as is H$\alpha$
  (see Section \ref{Halpha} for more information on this subject).

\begin{figure}
\begin{center}
\includegraphics[width=9cm]{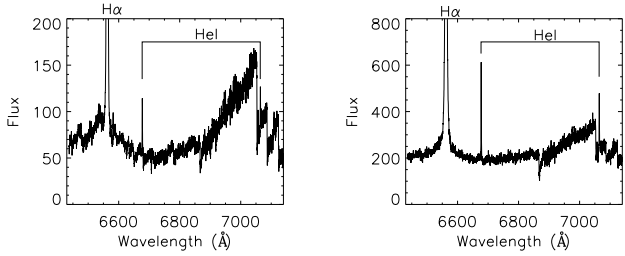}
\caption{Sample of spectra (\#82 on left panel and \#7 on right panel)
  with strong H$\alpha$ (6562.8 \AA{}), and He {\sc i} (6677.6 and
  7064.6 \AA{}) emission lines.}
\label{FigHeI}
\end{center}
\end{figure}

\item Four spectra (objects \#50, 59, 78 and 88) show the [N{\sc ii}]
  doublet at 6548.4 and 6583.4 \AA{} and the [S{\sc ii}] doublet at
  6715.8 and 6729.8 \AA{} (see examples in Figure \ref{FigHH}). These
  lines are indicative of ionised gas. Par-Lup3-4 (\#50) and Sz102
  (\#78) are related to Herbig-Haro (HH) objects HH600 and HH228
  respectively \citep{WH09}. They classified these objects to be HHs
  with the [S{\sc ii}] doublet line at 6717/6731 \AA{}. For 
  Par-Lup3-4 only these emission lines were detected. This is consistent 
  with \citet{Hue10} who show that this source has a close to edge-on disk. 
  This would make the source appear underluminous.

  The two other
  objects that show similar spectra, SSTc2dJ160708.6-391407 (\#59) and
  Sz133 (\#88), are not yet classified as Herbig-Haro, and are thus
  designated as Herbig-Haro candidates.

\begin{figure}
\begin{center}
\includegraphics[width=9cm]{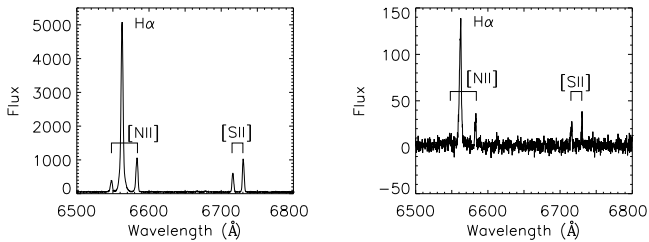}
\caption{Sample of spectra (\#78 on left panel and \#59 on right
  panel) with [N{\sc ii}] and [S{\sc ii}] doublets (at 6548.4 and
  6583.4, and 6715.8 and 6729.8 \AA{} respectively). Those sources may
  be associated with Herbig-Haro objects.}
\label{FigHH}
\end{center}
\end{figure}

\item One spectrum, IRAS15589-4132 (\#14), shows lines at 6757.95, 6774,
  6794.8, 6931.5 and 6947.8 \AA{}, and no other features (see Figure
  \ref{FigIRA}). These lines are identified as the doublets [N {\sc
      ii}] (at 6548.4 and 6583.4 \AA{}) and [S {\sc ii}] (at 6715.8
  and 6729.8 \AA{}) and the strongest H$\alpha$ line (at 6562.8 \AA{})
  at a redshift $z=0.032$. This suggests that the
  object is a red-shifted galaxy rather than a YSO in the Lupus
  Cloud. In the NASA/IPAC Extragalactic Database (NED), there is a
  galaxy classified at the position of this object:
  2MASSJ16022165-4140536. For further analysis, this object will no
  longer be considered part of the sample, leaving 91 sources in
  total.

\begin{figure}
\begin{center}
\includegraphics[width=9cm]{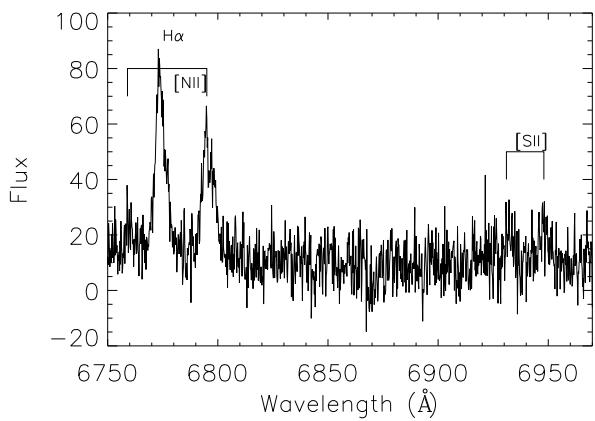}
\caption{Spectrum of IRAS15589-4132. The emission lines correspond
  with the spectral lines [N {\sc ii}] (at 6548.4 and 6583.4 \AA{})
  and [S {\sc ii}] (at 6715.8 and 6729.8 \AA{}). This object is
  identified as a galaxy at redshift $z=0.0324$.}
\label{FigIRA}
\end{center}
\end{figure}

\end{itemize}

\subsection{Spectral Types}\label{spec_feat}

\begin{figure}
\begin{center}
\includegraphics[width=9cm]{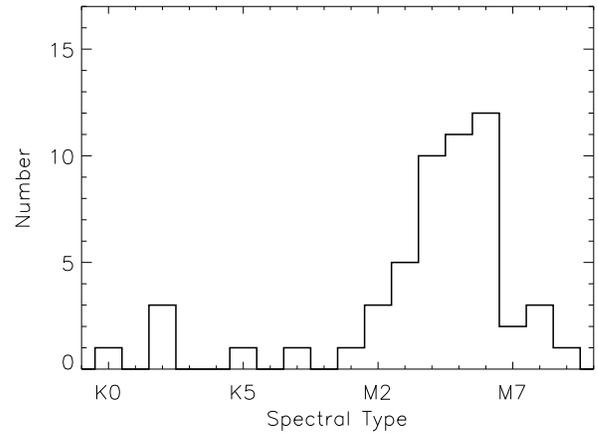}
\caption{Distribution of spectral types found in this work. The
  distribution peaks at M4--M6.}
\label{FigSpt}
\end{center}
\end{figure}

Following the method described above, the resulting spectral types
with their error ranges are listed in Table \ref{TabSpTAV}. The
objects are mostly M-type stars, but $10\%$ are K-type. The histogram
in Figure \ref{FigSpt} shows the spectral type distribution, peaking
at M6. 

40 of these 54 objects were designated by \citet{Merin08} as
YSOs. Spectral classification is found in the literature for 31 of the
54 objects ($57\%$). The literature classifications match well (within
one sub-class) with those derived here. One object, however, differs:
object \# 26 is classified by \citet{Lopez05} as M5, but as K0 (range
from K0 to K2) here. Their spectral type determination of M5 was 
based on photometry, which is not as reliable as the spectral classification 
performed here. For further analysis, the values derived in this work are
used.

\subsection{Effective Temperature}\label{teff}

Because spectral types are directly related to effective temperature
$T_{eff}$, the temperatures of the stars classified here can be
obtained (see Table \ref{TabAM}). For spectral types in the range K --
M0, the relationship adopted by \citet{Kenyon95} was used, while the
relationship from \citet{Luhman03} was adopted for types later than
M0. The uncertainties in the temperature are determined directly from
the spectral type uncertainties.

\section{Spectral Energy Distributions}\label{SED}

For a given star with known spectral type (and therefore $T_{eff}$),
an SED can be constructed using the correct stellar atmosphere model
and additional photometric data. The templates for stellar atmospheres
used here are the NextGen Model Atmospheres \citep{Haus99,Allard00}.
In addition, various photometric datasets are available for the
objects in Lupus. Optical and near-IR photometry in the
R$_{\mathrm{C}}$, I$_{\mathrm{C}}$ and Z$_{\mathrm{WFI}}$ bands (at
0.6517, 0.7838 and 0.9648 $\mu$m, respectively) was obtained from
Comer\'on et al. (2010, private communication). The data were taken
with the Wide Field Imager (WFI) at the La Silla 2.2 m Telescope in
Chile. Near-infrared Two Micron All-Sky Survey (2MASS) photometry at
J, H and K (at 1.235, 1.662 and 2.159 $\mu$m, respectively) is
publicly available. Finally, {\it Spitzer} mid-infrared data at the
IRAC and MIPS bands (at 3.6, 4.5, 5.8, 8.0, 24 and 70 $\mu$m,
respectively) are also available
online \footnote{htpp://ssc.spitzer.caltech.edu/spitzerdataarchives/}
\citep{Evans03}.

For each object, its photometric data and corresponding NextGen model
atmosphere are used (matching its spectral type and the gravitational
acceleration for a pre-main-sequence star). The observed photometric
data are corrected for extinction at all wavelengths from the object's
visual extinction ($A_V$), using the extinction law by \citet{WD01}
with $R_V =$ 5.5. This extinction is determined by SED fitting, 
with considering the derived spectral types. 
Subsequently, the NextGen atmosphere is scaled to
the extinction corrected photometric data at a given band. The scaling
band used is either the z$_{\mathrm{WFI}}$ or $J$ band, depending on
the availability of the photometry and the quality of the fit. In
Table \ref{TabSpTAV}, the extinction values and normalization bands
are presented for all fitted objects.

\begin{figure*}
\begin{center}
\includegraphics[width=16cm]{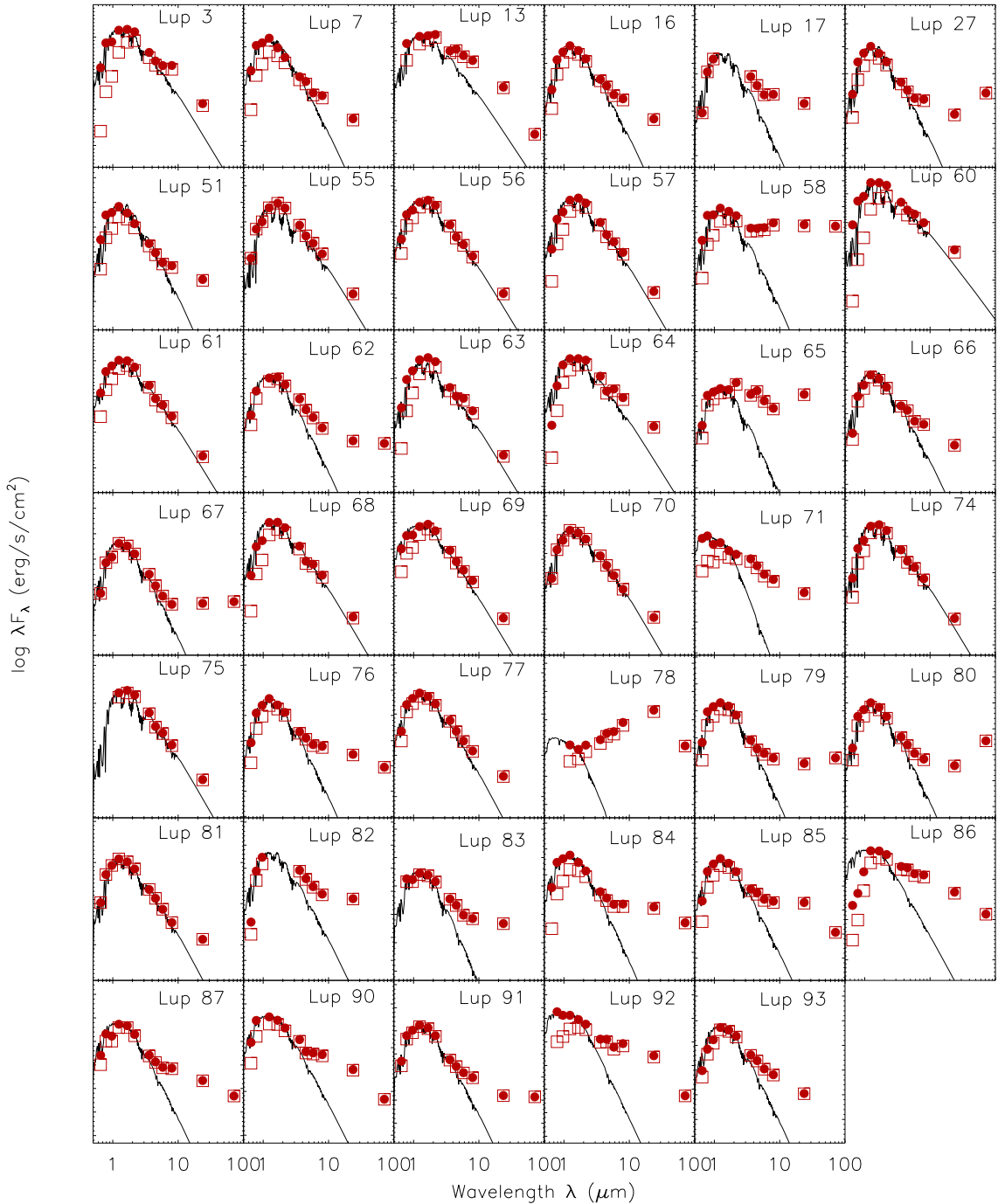}
\caption{SEDs of all the objects. The black curve is the model
  atmosphere of a star at its given temperature. The open points are
  the observed optical, 2MASS and \textit{Spitzer} data. The closed
  points are the extinction corrected data. Notice the wide variety in
  SEDs (and thus disk shapes).}
\label{FigSED}
\end{center}
\end{figure*}

Figure \ref{FigSED} shows the constructed SEDs for 41 out of 54
objects. Here, the solid black line is the NextGen model atmosphere,
the open squares are the observed photometry while filled circles show
the extinction corrected data. SEDs could not be produced for 13 out
of the 54 spectrally classified objects. Either not enough infrared (both
2MASS as Spitzer) photometry is available, or the different
photometric data (optical, 2MASS and Spitzer) did not match well with
each other. Eight of them are in crowded regions (18, 19, 23, 26, 47, 
48, 54, 89), which may have contaminated the photometry. None of the 13 
objects are known variable stars.

A wide variety of SEDs can be seen in Figure \ref{FigSED}.
Consequently, a wide variety in disk types is inferred. The infrared
spectral index slope can be used as a characteristic to distinguish
between Class I, II and III objects, as defined by Lada. For the 39
objects with available data in K and MIPS1, the slope can be
calculated as:

\[
\alpha_{2-24\mu m}=\frac{(\lambda F_{\lambda})_{24\mu m}-(\lambda F_{\lambda})_{2\mu m}}{\lambda_{24\mu m}-\lambda_{2\mu m}}
\]

These values are shown in Table \ref{TabSpTAV}. One object has
$\alpha_{2-24\mu m}$ consistent with a Class I source ($\alpha > 0$)
and 15 consistent with Class III sources ($\alpha < -2$), the majority
of them are Class II, disk sources. Additionally, object \#67 is a
confirmed cold disk, as studied by \citet{Merin10}. Cold disks are
disks with large inner dust holes (i.e. lack of warm dust). Objects \#
27, 79 and 80 have SEDs consistent with inner dust holes. Although
confirmation of their nature as cold disks requires SED modelling that
is out of the scope of this work, 4 out of 41 objects with SEDs in
Figure \ref{FigSED} amount to $\sim$10\% of cold disks in the sample,
which is consistent with the range of 4 -- 12\% of cold disk frequency
derived by \citet{Merin10}.

\subsection{Luminosities}\label{lum}

Stellar and disk luminosities can be derived for the objects for which
SEDs could be plotted. Stellar luminosity is calculated with the
formula:

\[
L_{\ast} = 4 \pi D^2 \int F_{\lambda} d\lambda
\]

\noindent where $D$ is the distance to the source and $F_{\lambda}$ is
the stellar flux -- an integration of the NextGen model atmosphere
scaled and normalized to the dereddened photometric data. Errors are
derived from the uncertainty in the distance and flux, which includes
an error in $A_V$ of $\pm 0.5$ mag. The disk luminosity is the
integrated excess emission above the stellar photosphere. Errors are 
derived from the stellar luminosity errors. Those two
values are shown in Table \ref{TabAM}.\\

\section{H-R diagram}\label{HR}

The age and mass of a star can be derived from theoretical tracks,
overlaid on physical H-R diagrams. The position of an object in the
H-R diagram is determined by its temperature (derived in Section
\ref{teff}) and luminosity (derived in Section \ref{lum}).

\subsection{Results}

\begin{figure}
\begin{center}
\includegraphics[width=8.2cm]{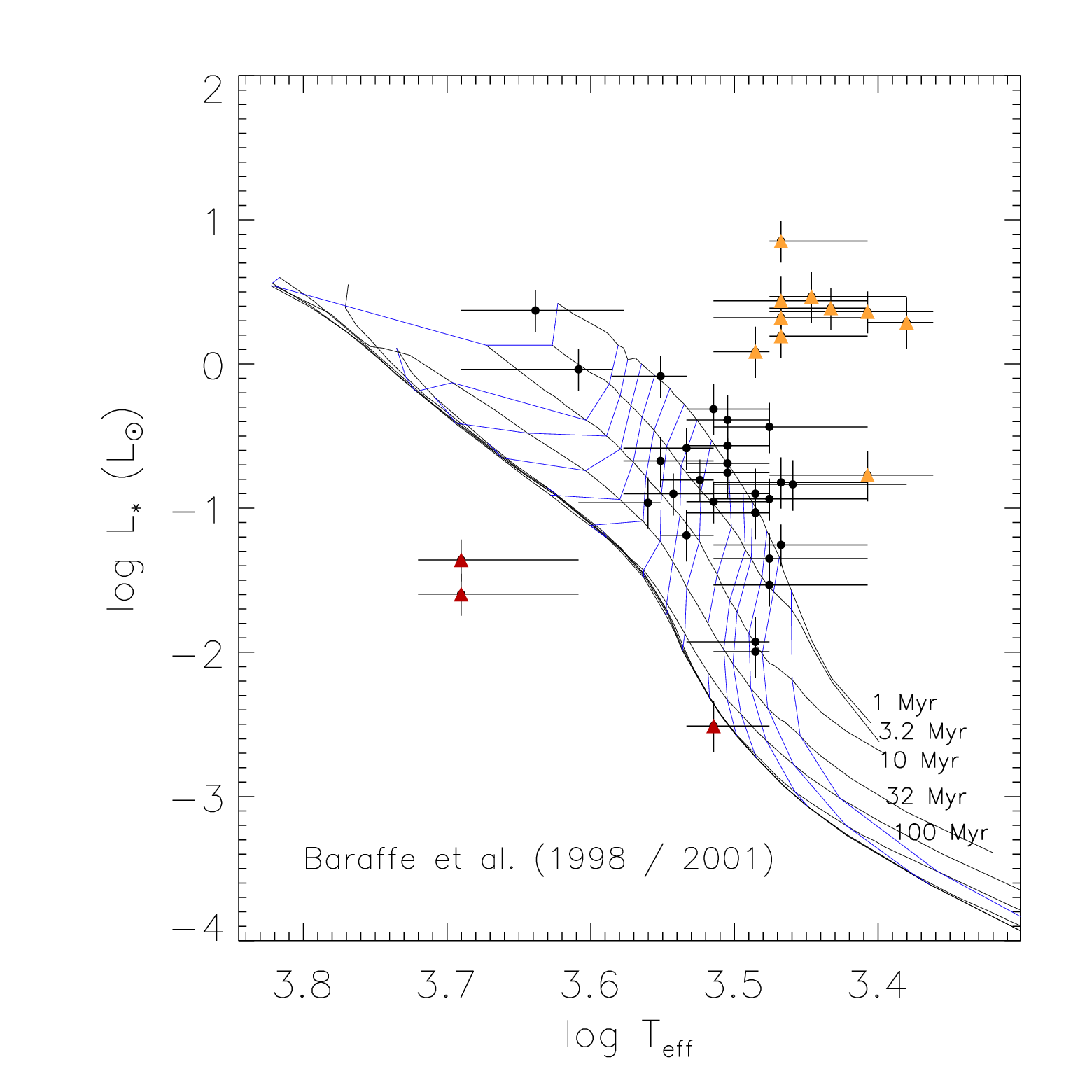}
\includegraphics[width=8.2cm]{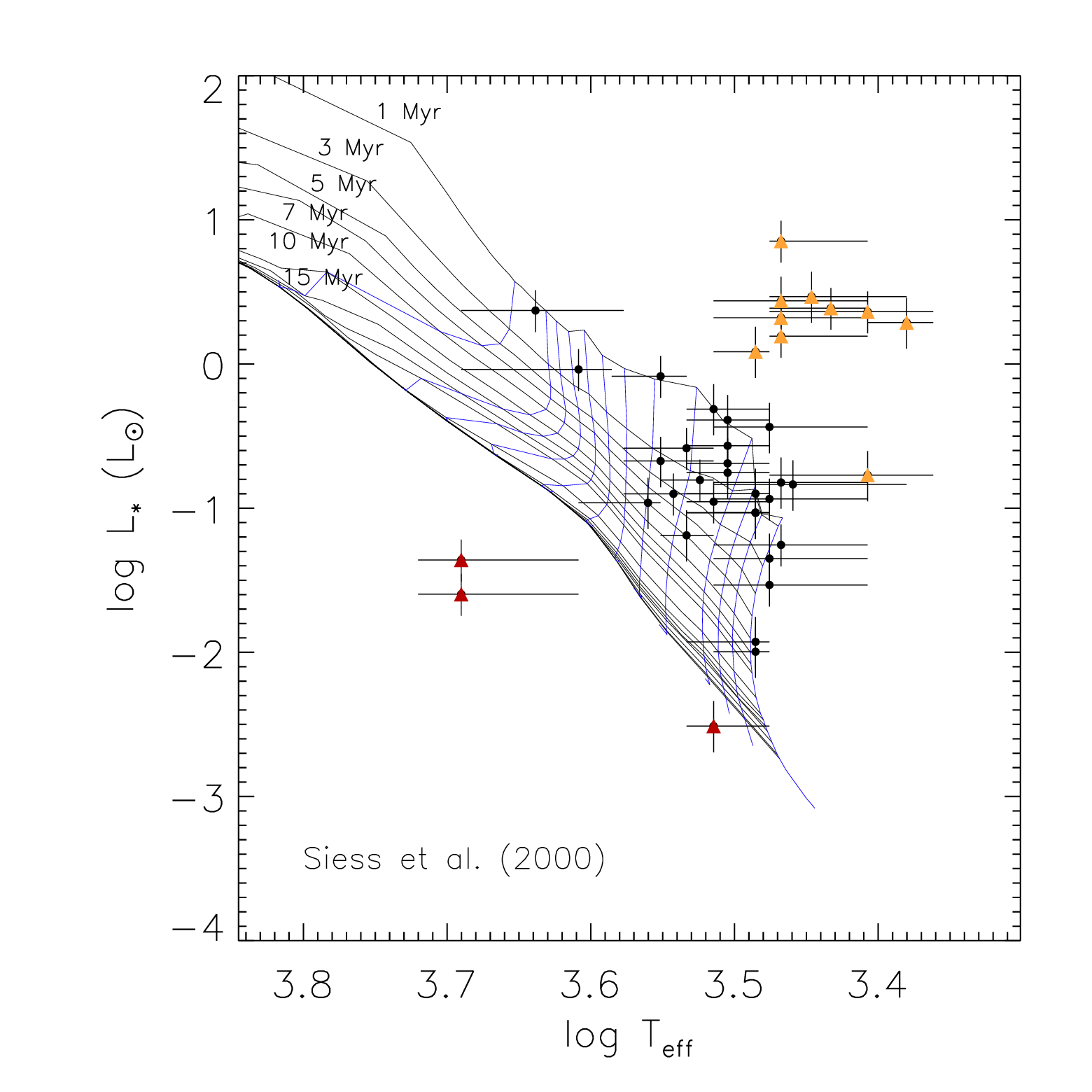}
\caption{HR diagram of the objects. Overlaid are the isochrones
  (black) and mass tracks (blue). The top panel shows the tracks from
  \citet{Baraffe98,Baraffe01} and the bottom panel the tracks from
  \citet{Siess00}. The corresponding ages (in Myr) are marked. 
  The red and yellow circles are objects that
  do not follow the tracks.}
\label{FigHR}
\end{center}
\end{figure}

The pre-main sequence evolutionary tracks of
\citet{Baraffe98,Baraffe01} and \citet{Siess00} are overlaid in the
H-R diagrams where the objects in this sample are placed (Figure
\ref{FigHR}). Thirteen objects are outside the range of the tracks
(red and yellow circles in Figure \ref{FigHR}).

Three objects (marked in red) are below the tracks: ID \# 65, 71 and
78. By looking at their SEDs in Figure \ref{FigSED}, it can be seen
that those objects 
have substantial IR excess. The stellar fluxes
are also not that high. These may be objects still surrounded by
envelopes, which could be responsible for their low
luminosity. Another explanation is that the objects are seen edge-on,
such that the dust in the disk blocks some of the light of the
star. In both cases, the calculated luminosity will be too low,
resulting in an incorrect placement in the HR diagram. 
Yet another reason may be that these are background sources and are in 
fact at a larger distance, which would increase the calculated luminosity.

Ten objects (marked in yellow) are above the tracks: ID 3, 13, 55, 60,
61, 63, 64, 68, 74 and 75.  
An easy explanation for the misplacement would be that the
assumed distance is wrong and that the objects are actually closer to
us. However, even with very small distances (e.g. 60 pc), the objects
are still above the tracks. Other effects have to play a role. 
None of them have detected lithium 
absorption and none of them, except object 13, show any infrared excess.
One option is that they are part of a binary system that is unresolved,
resulting in a stellar luminosity that is too high. Another option is
that they are evolved AGB-stars, which would explain the IR excess seen in 
object 13. The working assumption is that these
objects, however, do not belong to the Lupus Clouds.

\begin{figure}
\begin{center}
\includegraphics[angle=270,width=8.2cm]{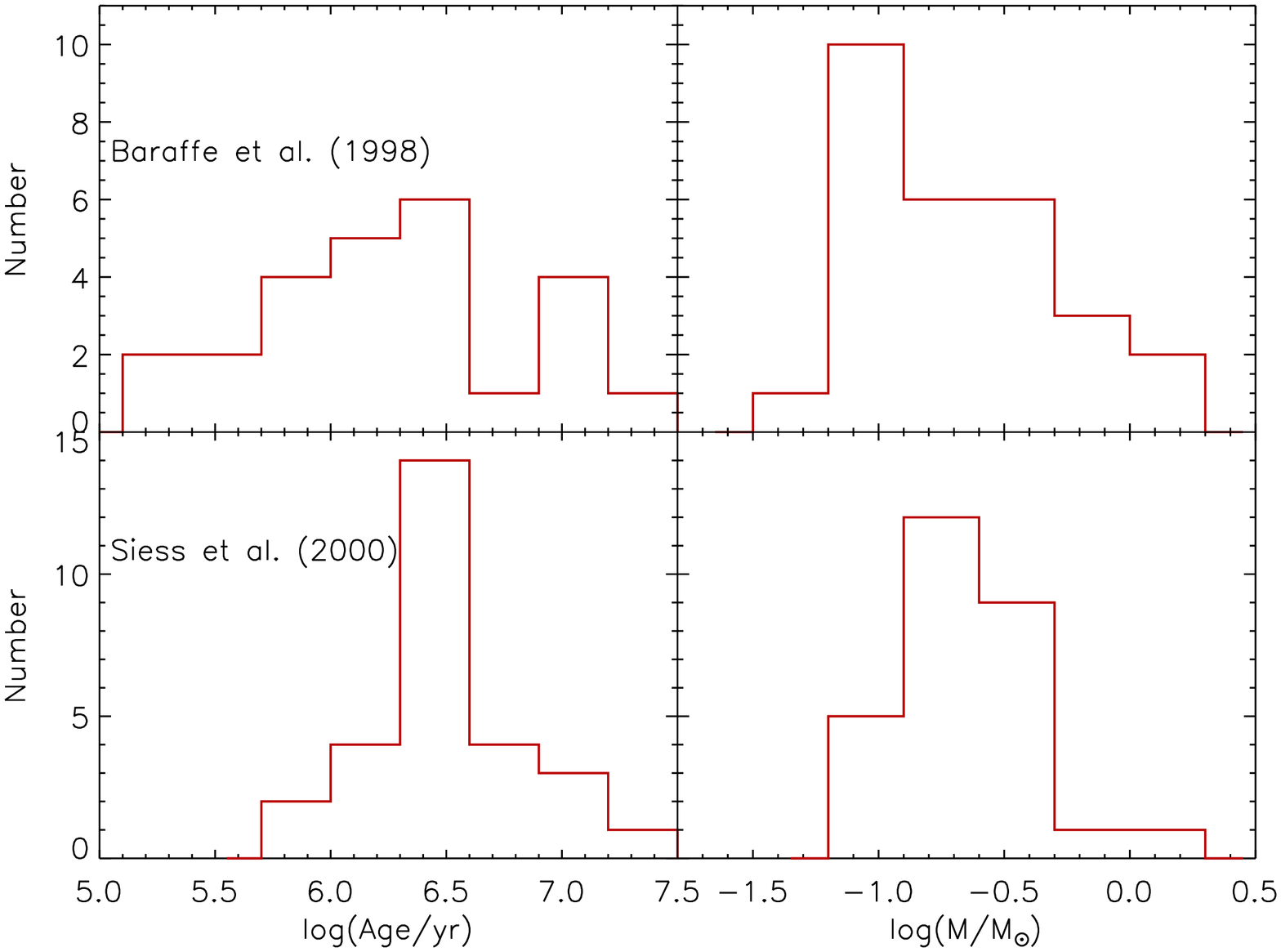}
\caption{Histograms of the ages and masses as derived from the two
  models.}
\label{FigAM}
\end{center}
\end{figure}

For those 13 objects, no ages and masses were derived. For the other
objects, the effective temperature, stellar and disk luminosity, age
and mass can be found Table \ref{TabAM}. As can be seen from Figure
\ref{FigAM}, the two models agree on the range of ages and masses,
albeit with a different distribution. The mean age is 3.6 Myr for the
Baraffe models and 4.4 Myr for the Siess models. For the masses, both
models determine a mean stellar mass of only 0.3 M$_{\odot}$. This
result is in agreement with that derived by \citet{Merin08}. When
deriving a luminosity function for the young stellar population in
Lupus, the authors found it to peak at 0.2 L$_{\odot}$, which
corresponds to 0.2 M$_{\odot}$ according to the PMS evolutionary
tracks of \citet{Baraffe98}.

\section{Accretion based on H$\alpha$ emission}\label{Halpha}

From the data presented here, 45 of the objects (48 $\%$) show the
H$\alpha$ line in emission. The strength of the H$\alpha$ emission
line is often used to distinguish between two classes of pre-MS stars:
classical T Tauri stars (CTTS) and weak-line T Tauri stars
(WTTS). CTTS are believed to correspond to stars that are actively
accreting and WTTS are not, respectively. To this end, different
methods have been proposed in the literature. Here, two methods are
explored: the equivalent width and the full width of H$\alpha$ at
$10\%$ of its peak intensity.

\subsection{H$\alpha$ Equivalent Width}

The value of equivalent width (EW) can be used to distinguish between
accreting and non-accreting objects, with the dividing threshold
depending on the spectral type of the star. \citet{WB03} proposed that
a star is a CTTS if EW[H$\alpha$] $\geq$ 3 \AA{} for K0-K5 stars,
EW[H$\alpha$] $\geq$ 10 \AA{} for K7-M2.5 stars, EW[H$\alpha$] $\geq$
20 \AA{} for M3-M5.5 stars and EW[H$\alpha$] $\geq$ 40 \AA{} for
M6-M7.5 stars. \citet{Bar03} extended this criterion to the 
substellar regime. Positive values designate emission. It is 
important to note that stars with EW[H$\alpha$]
values lower than these proposed levels are not necessarily
WTTS. Extra confirmation is needed from other diagnostics such as the
Lithium abundance.

Following the criterion of \citet{WB03}, 24 of the 45 objects 
(53.3\%) are considered
to be actively accreting (values can be found in Table
\ref{TabHa}). For objects \# 22, 49, 50 and 59, no spectral type is
available. However, since the EWs of these 4 objects are greater than
40 \AA{}, they are considered accretors anyway.

\subsection{Full width of H$\alpha$ at $10\%$ of peak intensity}

\citet{WB03} proposed the use of another quantity to distinguish
accreting and non-accreting objects: the full width of the line at
$10\%$ of the peak intensity, H$\alpha [10\%]$. Lines with H$\alpha
[10\%] > 270$ km s$^{-1}$ are, according to their measurements,
assumed to be accreting. Based on physical reasoning as well as
empirical findings, \citet{Jaya03} adopted 200 km s$^{-1}$ as a more
reasonable accretion cut-off, especially for low-mass stars. This is
the cut-off value used in this work.

\begin{figure*}
\begin{center}
\includegraphics[width=15cm]{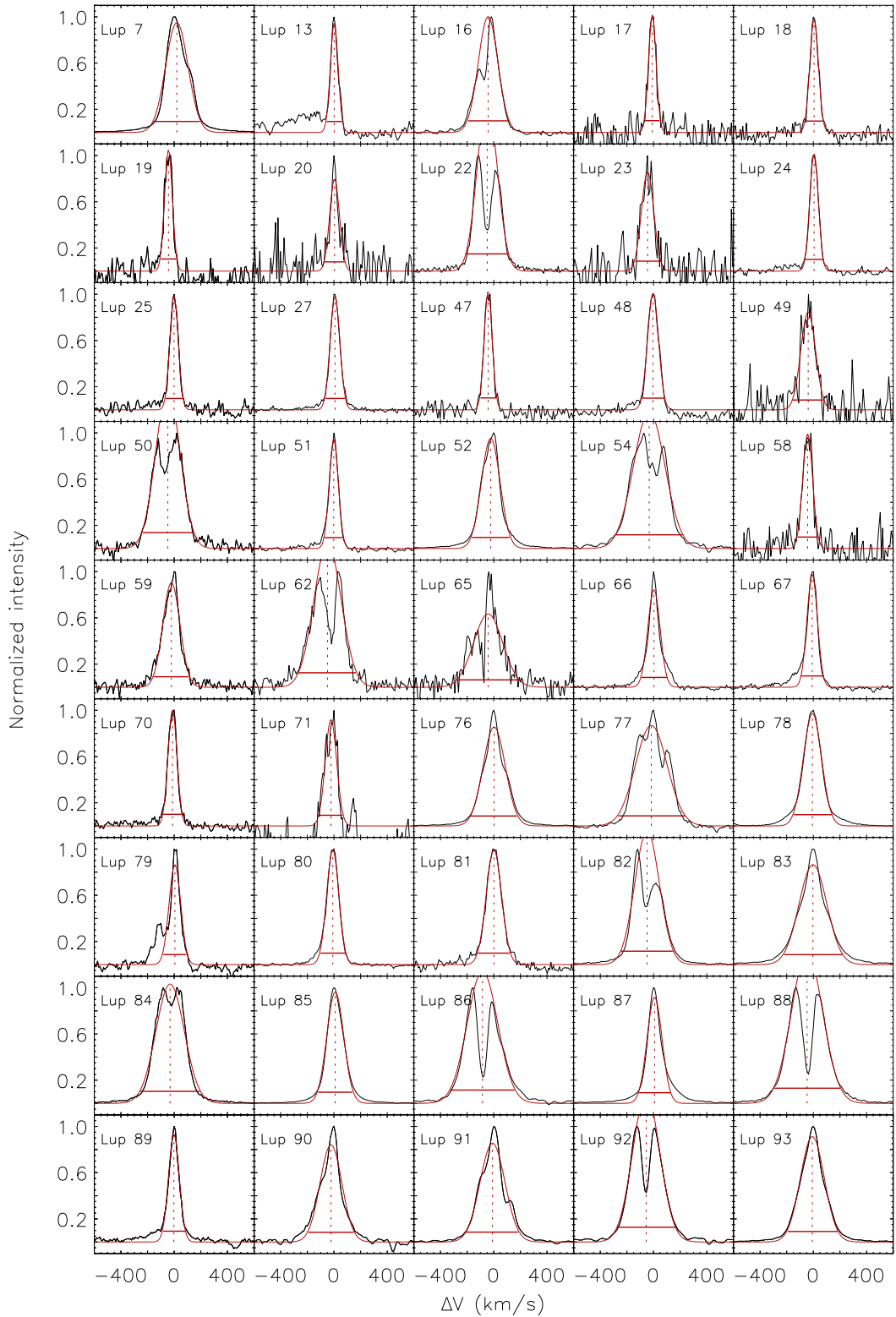}
\caption{Continuum subtracted, normalized profiles of the H$\alpha$
  emission lines (black line) overplotted with a Gaussian fit (red
  solid line). The red dotted line marks the centre of the Gaussian
  fit and the horizontal red line marks the width at 10\% intensity. }
\label{FigGauss}
\end{center}
\end{figure*}

By performing a Gaussian fit (see Figure \ref{FigGauss}), H$\alpha
[10\%]$ values can be obtained. To account for the resolution, the
measured FWHM of each profile must be deconvolved assuming a Gaussian
instrumental profile. For this sample, the H$\alpha [10\%]$ values can
be found in Table \ref{TabHa}. The errors are propagated from the
error in the Gaussian fit to the error on the spectral resolution.
According to the criterion of \citet{Jaya03}, 25 of the 45 objects
(55.5\%) are considered accretors. The other 20 objects are not
consistent with CTTS, even within their uncertainties. They are
classified as non-accretors.

\begin{figure}
\begin {center}
\includegraphics[width=6cm,angle=270]{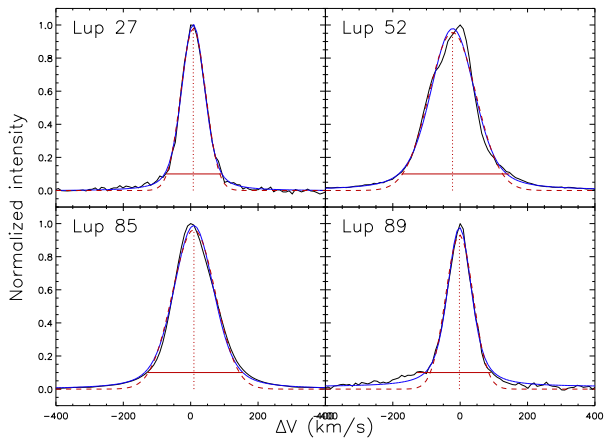}
\caption{Sample of continuum subtracted, normalized profiles of the
  H$\alpha$ emission lines overplotted with a Gaussian fit (red dashed
  line) and a Voigt fit (thick blue solid line). The red dotted line marks
  the center of the Gaussian fit.}
\label{FigVoi}
\end{center}
\end{figure}

Some line profiles show broad wings at the bottom, which cannot be
properly fitted by a Gaussian. In order to reproduce those lines,
Voigt profiles were fitted to all the lines. In most cases, the fit is
either the same or slightly better at the wings as can be see in
Figure \ref{FigVoi}. However, this correction has no significant
effect on the full width at $10\%$. Due to this insignificant
difference, only the values from the Gaussian fits are further
considered in this work.

The EW and H$\alpha [10\%]$ methods are generally consistent in the
classification of sources. Only for three objects (\#65, 80, 81) do
the two methods give discrepant results. This may be the result of 
the different cut-offs in the EW method, depending on the spectral type.

\subsection{Mass accretion rate}

\begin{figure}
\begin{center}
\includegraphics[width=6cm,angle=270]{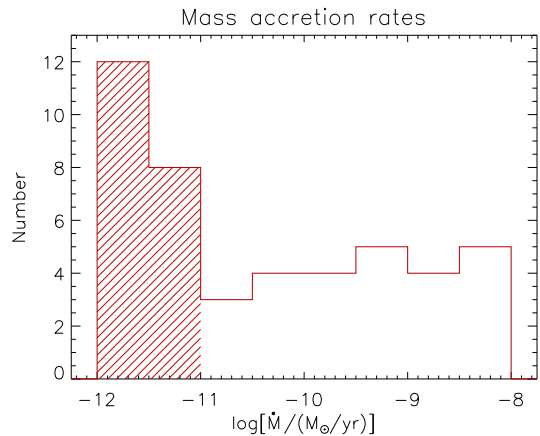}
\caption{Histogram of the mass accretion rates derived from the
  H$\alpha[10\%]$ value. The shaded area are objects out of the range
  for which the relationship of \citet{Natta04} was calibrated. Those
  objects are generally classified as non-accretors.}
\label{HistHa}
\end{center}
\end{figure}

The H$\alpha [10\%]$ value can not only be used as an indicator of
accretion, but also allows a quantitative estimate of the mass
accretion rate. \citet{Natta04} derived a correlation between H$\alpha
[10\%]$ and mass accretion rate $\dot{M}_{ac}$ (derived from
veiling and H$\alpha$ profile model fittings):

\[
\log\dot{M_{ac}}= -12.89 (\pm 0.3) + 9.7 (\pm 0.7) \cdot 10^{-3} \mathrm{H}\alpha[10\%]
\]

\noindent where $\dot{M}_{ac}$ is in M$_{\odot}$ yr$^{-1}$ and
H$\alpha [10\%]$ in km s$^{-1}$. The relation derived by
\citet{Natta04} was only calibrated for $-11 < \log\dot{M} < -6$. The
cut-off of 200 km s$^{-1}$ agrees with a cut-off of $10^{-10.95}$
M$_{\odot}$ yr$^{-1}$ for $\dot{M}$. With this relation, mass
accretion rates can be calculated for the objects in this sample with
H$\alpha$ in emission (see Table \ref{TabHa}). In Figure \ref{HistHa}
the distribution of the mass accretion rates is shown. Typical
uncertainties are an order of magnitude. Most of the accretors have
mass accretion rates that are consistent with the typical range for
classical T Tauri stars \citep{Natta04,Sicilia10}.

A fit of the mass accretion rate as a function of the stellar mass is 
attempted. This fit reveals that a $\dot{M}_{ac} \propto M^2$ relation is 
valid for both models, within the large errors. The same relation, 
and huge scatter, can be found in the literature \citep{Natta05,Alcala11}.

\section{Discussion}\label{disc}

\noindent Our work can be added to some of the objects classified as
young stars by \citet{Merin08} that have previously been spectrally
classified in the literature in different studies
\citep{Lopez05,Allers06,Allen07,Merin07,Comeron08}. This amounts to 91
of the 159 IR excess sources in the sample of \citet{Merin08} to have
known spectral types. This fairly complete sample can be compared to
other nearby star-forming regions with equally complete stellar
characterization.

\begin{figure}
\begin{center}
\includegraphics[width=9cm]{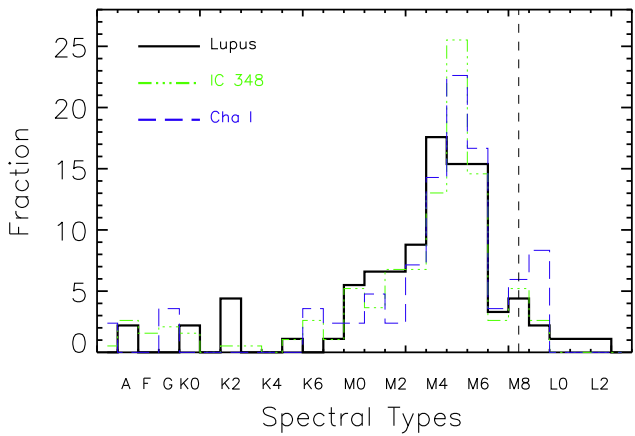}
\caption{Distribution of spectral types in Lupus (black solid line),
  IC 348 (green dot-dashed line) and Chamaeleon I (blue dashed line).
  For Lupus, the distribution includes
  both our results as well as literature spectral types for sources
  not in our sample.}
\label{FigSptComp}
\end{center}
\end{figure}

\citet{Luhman03} studied the young cluster IC348. Spectral types are
determined with low-resolution optical spectroscopy for 169
objects. Similarly, \citet{Luhman07} determined spectral types for the
85 targets in the star forming region Chamaeleon I. Figure
\ref{FigSptComp} shows the distribution of spectral types in Lupus
(this work + literature), in IC 348 and Chamaeleon I. The vertical
dashed line shows the completeness of the IC 348 and Cha I samples,
which is comparable to the limit achieved by \citet{Merin08}. It can
be seen that the three regions show very similar distributions of
stellar types, from which is inferred that the IMF of Lupus is very
similar to those of IC 348 and Cha I.

It is important to note that the sample of \citet{Merin08} is selected
on the basis of IR excess and it is, by definition, biased against
young stars without disks (i.e. no IR excess). A similar study
characterizing the young stellar population without disks in Lupus is
still lacking.

When comparing the stellar luminosity with the disk luminosity, a 
correlation $L_{disk}\propto L_{\ast}$ can be found with a correlation 
coefficient of $0.82$. The 
spectral index $\alpha_{2-24\mu m}$ is also compared with the stellar 
parameters (mass, age, luminosity). Slight trends can be seen in 
that the IR excess increases with mass and decreases with the 
stellar and disk luminosity. However, the scatter is huge in all the 
fits. More observations and a larger sample are needed to make significant 
conclusions.

\section{Conclusions}\label{conclus}

\noindent We have presented a spectroscopic survey at optical
wavelengths designed to determine spectral types and confirm the
pre-main sequence nature of a sample of young stellar objects in the
Lupus Clouds found in the {\it Spitzer} c2d survey.

\begin{itemize}

\item Spectral types were determined for 54 stars belonging to Lupus
  I, III or IV. The sample consists mostly of M-type stars (90\%), but
  also a few K-type stars. No early type object was found in this
  sample. The distribution of spectral types peaks at M4--M6. The
  distribution of spectral types is very similar to that of well
  studied star-forming regions like IC 348 and Chamaeleon I.

\item The objects were placed in a H-R diagram after effective
  temperatures and luminosities were derived. Comparison with
  theoretical isochrones and mass tracks from models of
  \citet{Baraffe98} and \citet{Siess00} yield individual ages and
  masses for the objects. 10 objects are too luminous to belong to
  the clouds, while 3 objects are too faint. The very faint objects
  could still belong to the clouds by having a remnant envelope or
  an edge-on disk dimming some of the stellar luminosity.

\item The theoretical models by \citet{Baraffe98} and \citet{Siess00}
  imply a population of YSOs concentrated in the age range between 1
  and 5 Myr. The mean age is found to be 3.6 and 4.4 Myr with the
  \citet{Baraffe98} and \citet{Siess00} tracks, respectively.
  Individual masses range from 0.1 to 1.0 M$_\odot$, with mean
  values of 0.3 M$_\odot$ for both models.

\item About half of the sample shows the H$\alpha$ line in emission,
  an important indicator of accretion. This relationship was explored
  in two different ways: the equivalent width of H$\alpha$, and its
  full width at 10\% of peak intensity. This confirms 25 objects (or
  56\% of the YSO sample) to be actively accreting objects classified
  as classical T Tauri stars. The quantitative estimate of the mass
  accretion rate $\dot{M}_{{\rm ac}}$ based on the full width of
  H$\alpha$ at 10\% of the peak intensity yields a broad distribution
  of values ($\sim10^{-11} - 10^{-8}$ M$_\odot$yr$^{-1}$), typical of
  T Tauri stars.

\end{itemize}

\section*{Acknowledgments}
Astrochemistry at Leiden is supported by a Spinoza grant from the
Netherlands Organization for Scientific Research (NWO) and by the
Netherlands Research School for Astronomy (NOVA) grants. This research
was based on observations made with ESO Telescopes at the Paranal
Observatory under programme ID 080C.0473-A. It also has made use
of the SIMBAD database, operated at CDS, Strasbourg, France. The
authors would like to acknowledge Greg Herczeg for sharing his low
resolution library of standards and for many useful discussions, Bruno
Mer\'{\i}n and Klaus Pontoppidan for comments on earlier versions of
the manuscript, and Loredana Spezzi \& Fernando Comer\'on for
providing the optical photometry.

\onecolumn

\begin{longtable}{cccllccc}
\caption{ID, name, cloud, position, observing dates and exposure times of the objects.}
\label{TabID}\\

\noalign{\smallskip} \hline \hline \noalign{\smallskip}
ID & Name & Cloud & RA & Dec & Observing Date & t$_{exp}$ & Mark\footnotemark[1] \\[0.5ex]
 & & & (deg) & (deg) &  & (s) & \\[0.5ex]
\hline\\[0.5ex]\endfirsthead

\multicolumn{8}{c}{{\tablename} \thetable{} -- Continued} \\
\noalign{\smallskip} \hline \hline \noalign{\smallskip}
ID & Name & Cloud & RA & Dec & Observing Date & t$_{exp}$ & Mark\footnotemark[1] \\[0.5ex]
 & & & (deg) & (deg) &  & (s) & \\[0.5ex]
\hline\\[0.5ex]\endhead

\endfoot

\hline\endlastfoot

\input{Table_ID2.tex} \footnotetext[1]{G means good spectra, U
  non-detections, F featureless spectra, and O not useful spectra. See
  \S{} \ref{spec_feat} for more details. }
\end{longtable}

\begin{longtable}{ccccccc}

\caption{Spectral types, visual extinction (A$_V$), normalization
  bands and the spectral index, calculated between the 2MASS K-band (2
  $\mu$m) and the MIPS1 band (24 $\mu$m), for the classifiable
  objects. Objects for which SEDs could not be produced do not have
  all values. The last column indicates if Lithium was detected.}
\label{TabSpTAV}\\

\noalign{\smallskip} \hline \hline \noalign{\smallskip}
ID & SpT & Range & A$_V$ & Band & $\alpha_{2-24\mu m}$ & Lithium? \\[0.5ex]
 &  &  & (mag) &  &  \\[0.5ex]
\hline\\[0.5ex]\endfirsthead

\multicolumn{7}{c}{{\tablename} \thetable{} -- Continued} \\
\noalign{\smallskip} \hline \hline \noalign{\smallskip}
ID & SpT & Range & A$_V$ & Band & $\alpha_{2-24\mu m}$ & Lithium? \\[0.5ex]
 & & & (mag) & & \\[0.5ex]
\hline\\[0.5ex]\endhead

\hline
\multicolumn{7}{l}{\emph{Continued on next page}}
\endfoot

\hline\endlastfoot
\input{Table_spavslope.tex}\footnotetext{Sources without ages
  and masses do not follow the evolutionary tracks. }
\end{longtable}

\begin{longtable}{cccccccc}
\caption{Effective temperature T$_{\mathrm{eff}}$, stellar and disk
  luminosity, L$_{\ast}$ and L$_{\mathrm{disk}}$, age and mass for
  Baraffe et al. (1998 - 2001) and Siess et al. (2000).}
\label{TabAM}\\
\noalign{\smallskip} \hline \hline \noalign{\smallskip}
ID & T$_{\mathrm{eff}}$ & L$_{\ast}$ & L$_{\mathrm{disk}}$ & Age$_B$ & Mass$_B$ & Age$_S$ & Mass$_S$ \\
 & (K) & (L$_{\odot}$) & (L$_{\odot}$) & (Myr) & (M$_{\odot}$) & (Myr) & (M$_{\odot}$) \\
\hline
\vspace{0.2ex}\endfirsthead
\hline
ID & T$_{\mathrm{eff}}$ & L$_{\ast}$ & L$_{\mathrm{disk}}$ & Age$_B$ & Mass$_B$ & Age$_S$ & Mass$_S$ \\
 & (K) & (L$_{\odot}$) & (L$_{\odot}$) & (Myr) & (M$_{\odot}$) & (Myr) & (M$_{\odot}$) \\
\hline
\vspace{0.2ex}\endhead
\hline
\multicolumn{8}{l}{\emph{continued on next page}}
\endfoot
\noalign{\smallskip} \hline \hline \noalign{\smallskip}\endlastfoot
\input{Table_prop.tex}
\end{longtable}

\begin{longtable}{ccccccc}

\caption{H$\alpha$ line widths and mass accretion rates. A positive 
 EW value indicates emission.} \label{TabHa}\\

\noalign{\smallskip} \hline \hline \noalign{\smallskip}
ID & EW[H$\alpha$] & CTTS? & H$\alpha$ [$10\%$] & CTTS? & log $\dot{M}_{\mathrm{ac}}$ & Double peak? \\[0.5ex]
 & (\AA{}) &  & (km s$^{-1}$) &  & (M$_{\odot}$ yr$^{-1}$) &  \\[0.5ex]
\hline\\\endhead
\hline
\multicolumn{7}{l}{\emph{Continued on next page}}
\endfoot
\hline\endlastfoot

\input{Table_ha.tex}

\end{longtable}

\label{lastpage}

\end{document}

%% file: Table_ID2.tex
$1$ & 2MASSJ16075475-3915446 & Lupus III & $241.9780$ & $-39.2623$ & 2008-02-21
 & $630.0$ & 
U
 \\
$2$ & 2MASSJ16080175-3912316 & Lupus III & $242.0070$ & $-39.2087$ & 2008-02-21
 & $630.0$ & 
U
 \\
$3$ & 2MASSJ16080618-3912225 & Lupus III & $242.0260$ & $-39.2062$ & 2008-02-21
 & $630.0$ & G \\
$4$ & 2MASSJ16081497-3857145 & Lupus III & $242.0620$ & $-38.9541$ & 2008-02-21
 & $630.0$ & 
U
 \\
$5$ & 2MASSJ16084747-3905087 & Lupus III & $242.1980$ & $-39.0856$ & 2008-02-21
 & $630.0$ & 
U
 \\
$6$ & 2MASSJ16085373-3914367 & Lupus III & $242.2240$ & $-39.2436$ & 2008-02-21
 & $630.0$ & 
U
 \\
$7$ & 2MASSJ16085529-3848481 & Lupus III & $242.2300$ & $-38.8134$ & 2008-02-22
 & $1000.0$ & G \\
$8$ & 2MASSJ16085953-3856275 & Lupus III & $242.2480$ & $-38.9411$ & 2008-02-21
 & $630.0$ & 
U
 \\
$9$ & AKC2006-17 & Lupus I & $234.8640$ & $-34.8121$ & 2008-02-23 & $800.0$ &  U
 \\
$10$ & IRACJ16083010-3922592 & Lupus III & $242.1250$ & $-39.3831$ & 2008-02-26
 & $1000.0$ & 
U
 \\
$11$ & IRACJ16084679-3902074 & Lupus III & $242.1950$ & $-39.0354$ & 2008-02-21
 & $630.0$ & 
U
 \\
$12$ & IRACJ16093418-3915127 & Lupus III & $242.3920$ & $-39.2535$ & 2008-02-26
 & $1000.0$ & 
U
 \\
$13$ & IRAS15567-4141 & Lupus IV & $240.0310$ & $-41.8302$ & 2008-02-21 & $550.0
$ & G \\
$14$ & IRAS15589-4132 & Lupus IV & $240.5900$ & $-41.6816$ & 2008-02-24 & $
1000.0$ & O \\
$16$ & Lup604s & Lupus III & $242.0010$ & $-39.0497$ & 2008-02-21 & $630.0$ & G
 \\
$17$ & Lup607 & Lupus III & $242.1170$ & $-39.2194$ & 2008-02-21 & $630.0$ & G
 \\
$18$ & Lup608s & Lupus III & $242.2850$ & $-39.0619$ & 2008-02-21 & $630.0$ & G
 \\
$19$ & Lup617 & Lupus III & $242.2010$ & $-39.1554$ & 2008-02-21 & $630.0$ & G
 \\
$20$ & Lup650 & Lupus III & $242.4580$ & $-38.8174$ & 2008-02-22 & $1000.0$ & G
 \\
$21$ & Lup654 & Lupus III & $241.8480$ & $-39.0861$ & 2008-02-21 & $630.0$ & F
 \\
$22$ & Lup706 & Lupus III & $242.1560$ & $-39.3864$ & 2008-02-26 & $1000.0$ & F
 \\
$23$ & Lup710 & Lupus III & $242.3210$ & $-39.4527$ & 2008-02-23 & $1000.0$ & G
 \\
$24$ & Lup714 & Lupus III & $241.9950$ & $-39.4097$ & 2008-02-26 & $1000.0$ & G
 \\
$25$ & Lup802s & Lupus III & $242.9630$ & $-38.8514$ & 2008-02-26 & $1000.0$ & G
 \\
$26$ & Lup810s & Lupus III & $242.4770$ & $-39.2009$ & 2008-02-22 & $1000.0$ & G
 \\
$27$ & Lup818s & Lupus III & $242.4850$ & $-38.9975$ & 2008-02-22 & $1000.0$ & G
 \\
$28$ & Lupus3MMS & Lupus III & $242.3250$ & $-39.0815$ & 2008-02-24 & $10.0$ & F
 \\
 &  &  &  &  & 2008-02-26 & $1000.0$ &  \\
$29$ & NTO2000-0526.9-5630 & Lupus III & $242.1170$ & $-39.0735$ & 2008-02-21 & 
$630.0$ & 
U
 \\
$30$ & NTO2000-0532.1-5616 & Lupus III & $242.2230$ & $-39.0692$ & 2008-02-21 & 
$630.0$ & 
U
 \\
$31$ & NTO2000-0536.7-5943 & Lupus III & $242.2430$ & $-39.1265$ & 2008-02-24 & 
$1000.0$ & F \\
$32$ & NTO2000-0536.7-5956 & Lupus III & $242.2430$ & $-39.1304$ & 2008-02-24 & 
$10.0$ & F \\
 &  &  &  &  & 2008-02-26 & $1000.0$ &  \\
$33$ & NTO2000-0537.4-5653 & Lupus III & $242.2460$ & $-39.0794$ & 2008-02-21 & 
$630.0$ & 
U
 \\
$34$ & NTO2000-0540.9-5757 & Lupus III & $242.2600$ & $-39.0970$ & 2008-02-21 & 
$630.0$ & F \\
$35$ & NTO2000-0546.4-5934 & Lupus III & $242.2830$ & $-39.1241$ & 2008-02-24 & 
$10.0$ & F \\
 &  &  &  &  & 2008-02-26 & $1000.0$ &  \\
$36$ & NTO2000-0554.9-5651 & Lupus III & $242.3180$ & $-39.0788$ & 2008-02-24 & 
$1000.0$ & F \\
$37$ & NTO2000-0558.8-5610 & Lupus III & $242.3350$ & $-39.0671$ & 2008-02-24 & 
$10.0$ & F \\
 &  &  &  &  & 2008-02-26 & $1000.0$ &  \\
$38$ & NTO2000-0601.7-5616 & Lupus III & $242.3460$ & $-39.0687$ & 2008-02-24 & 
$10.0$ & F \\
 &  &  &  &  & 2008-02-26 & $1000.0$ &  \\
$39$ & NTO2000-0605.1-5606 & Lupus III & $242.3610$ & $-39.0660$ & 2008-02-26 & 
$1000.0$ & 
U
 \\
$40$ & NTO2000-0605.6-5437 & Lupus III & $242.3630$ & $-39.0412$ & 2008-02-21 & 
$630.0$ & 
U
 \\
$41$ & NTO2000-0614.0-5414 & Lupus III & $242.3980$ & $-39.0349$ & 2008-02-21 & 
$630.0$ & 
U
 \\
$42$ & NTO2000-0615.6-5616 & Lupus III & $242.4050$ & $-39.0686$ & 2008-02-24 & 
$1000.0$ & 
U
 \\
$43$ & NTO2000-0615.6-5953 & Lupus III & $242.4050$ & $-39.1291$ & 2008-02-24 & 
$1000.0$ & 
U
 \\
$44$ & NTO2000-0615.8-5734 & Lupus III & $242.4060$ & $-39.0905$ & 2008-02-24 & 
$1000.0$ & 
U
 \\
$45$ & NTO2000-0617.7-5641 & Lupus III & $242.4140$ & $-39.0755$ & 2008-02-24 & 
$1000.0$ & 
U
 \\
$46$ & NTO2000-0619.6-5414 & Lupus III & $242.4210$ & $-39.0349$ & 2008-02-21 & 
$630.0$ & 
U
 \\
$47$ & Par-Lup3-1 & Lupus III & $242.0670$ & $-39.0511$ & 2008-02-21 & $630.0$
 & G \\
$48$ & Par-Lup3-2 & Lupus III & $242.1490$ & $-39.0632$ & 2008-02-21 & $630.0$
 & G \\
$49$ & Par-Lup3-3 & Lupus III & $242.2060$ & $-39.0942$ & 2008-02-21 & $630.0$
 & F \\
$50$ & Par-Lup3-4 & Lupus III & $242.2140$ & $-39.0917$ & 2008-02-21 & $630.0$
 & F \\
$51$ & SST-Lup3-1 & Lupus III & $242.9990$ & $-38.3940$ & 2008-02-25 & $1000.0$
 & G \\
$52$ & SSTc2dJ155925.2-423507 & Lupus IV & $239.8550$ & $-42.5853$ & 2008-02-26
 & $1000.0$ & G \\
$53$ & SSTc2dJ160000.6-422158 & Lupus IV & $240.0020$ & $-42.3659$ & 2008-02-26
 & $1000.0$ & O \\
$54$ & SSTc2dJ160002.4-422216 & Lupus IV & $240.0100$ & $-42.3708$ & 2008-02-26
 & $1000.0$ & G \\
$55$ & SSTc2dJ160111.6-413730 & Lupus IV & $240.2980$ & $-41.6250$ & 2008-02-21
 & $550.0$ & G \\
$56$ & SSTc2dJ160143.3-413606 & Lupus IV & $240.4300$ & $-41.6016$ & 2008-02-21
 & $550.0$ & G \\
$57$ & SSTc2dJ160229.9-415111 & Lupus IV & $240.6250$ & $-41.8530$ & 2008-02-24
 & $1000.0$ & G \\
$58$ & SSTc2dJ160703.9-391112 & Lupus III & $241.7660$ & $-39.1865$ & 2008-02-21
 & $630.0$ & G \\
$59$ & SSTc2dJ160708.6-391407 & Lupus III & $241.7860$ & $-39.2354$ & 2008-02-21
 & $630.0$ & F \\
$60$ & SSTc2dJ160755.3-390718 & Lupus III & $241.9800$ & $-39.1215$ & 2008-02-21
 & $630.0$ & G \\
$61$ & SSTc2dJ160803.0-385229 & Lupus III & $242.0130$ & $-38.8749$ & 2008-02-24
 & $1000.0$ & G \\
$62$ & SSTc2dJ160901.4-392512 & Lupus III & $242.2560$ & $-39.4200$ & 2008-02-23
 & $1000.0$ & G \\
$63$ & SSTc2dJ160934.1-391342 & Lupus III & $242.3920$ & $-39.2284$ & 2008-02-21
 & $630.0$ & G \\
$64$ & SSTc2dJ161000.1-385401 & Lupus III & $242.5000$ & $-38.9001$ & 2008-02-22
 & $1000.0$ & G \\
$65$ & SSTc2dJ161013.1-384617 & Lupus III & $242.5540$ & $-38.7712$ & 2008-02-22
 & $1000.0$ & G \\
$66$ & SSTc2dJ161019.8-383607 & Lupus III & $242.5830$ & $-38.6018$ & 2008-02-23
 & $1000.0$ & G \\
$67$ & SSTc2dJ161029.6-392215 & Lupus III & $242.6230$ & $-39.3707$ & 2008-02-23
 & $1000.0$ & G \\
$68$ & SSTc2dJ161034.5-381450 & Lupus III & $242.6440$ & $-38.2473$ & 2008-02-25
 & $1000.0$ & G \\
$69$ & SSTc2dJ161131.9-381110 & Lupus III & $242.8830$ & $-38.1861$ & 2008-02-25
 & $1000.0$ & G \\
$70$ & SSTc2dJ161144.9-383245 & Lupus III & $242.9370$ & $-38.5458$ & 2008-02-23
 & $1000.0$ & G \\
$71$ & SSTc2dJ161148.7-381758 & Lupus III & $242.9530$ & $-38.2994$ & 2008-02-25
 & $1000.0$ & G \\
$72$ & SSTc2dJ161204.5-380959 & Lupus III & $243.0190$ & $-38.1663$ & 2008-02-25
 & $1000.0$ & 
U
 \\
$73$ & SSTc2dJ161211.2-383220 & Lupus III & $243.0470$ & $-38.5389$ & 2008-02-22
 & $1000.0$ & 
U
 \\
$74$ & SSTc2dJ161219.6-383742 & Lupus III & $243.0820$ & $-38.6283$ & 2008-02-22
 & $1000.0$ & G \\
$75$ & SSTc2dJ161251.7-384216 & Lupus III & $243.2150$ & $-38.7044$ & 2008-02-22
 & $1000.0$ & G \\
$76$ & Sz100 & Lupus III & $242.1070$ & $-39.1003$ & 2008-02-21 & $300.0$ & G \\
 &  &  &  &  & 2008-02-23 & $200.0$ &  \\
 &  &  &  &  & 2008-02-26 & $200.0$ &  \\
$77$ & Sz101 & Lupus III & $242.1180$ & $-39.0922$ & 2008-02-21 & $630.0$ & G \\
$78$ & Sz102 & Lupus III & $242.1240$ & $-39.0530$ & 2008-02-21 & $630.0$ & G \\
$79$ & Sz103 & Lupus III & $242.1260$ & $-39.1030$ & 2008-02-21 & $630.0$ & G \\
$80$ & Sz104 & Lupus III & $242.1280$ & $-39.0968$ & 2008-02-21 & $630.0$ & G \\
$81$ & Sz107 & Lupus III & $242.1740$ & $-39.0269$ & 2008-02-21 & $630.0$ & G \\
$82$ & Sz108B & Lupus III & $242.1790$ & $-39.1040$ & 2008-02-21 & $630.0$ & G
 \\
 &  &  &  &  & 2008-02-23 & $200.0$ &  \\
 &  &  &  &  & 2008-02-26 & $200.0$ &  \\
$83$ & Sz110 & Lupus III & $242.2150$ & $-39.0549$ & 2008-02-21 & $300.0$ & G \\
 &  &  &  &  & 2008-02-23 & $200.0$ &  \\
 &  &  &  &  & 2008-02-26 & $200.0$ &  \\
$84$ & Sz113 & Lupus III & $242.2410$ & $-39.0396$ & 2008-02-21 & $630.0$ & G \\
$85$ & Sz114 & Lupus III & $242.2580$ & $-39.0867$ & 2008-02-21 & $300.0$ & G \\
 &  &  &  &  & 2008-02-23 & $200.0$ &  \\
 &  &  &  &  & 2008-02-26 & $200.0$ &  \\
$86$ & Sz118 & Lupus III & $242.4530$ & $-39.1880$ & 2008-02-22 & $1000.0$ & G
 \\
$87$ & Sz130 & Lupus IV & $240.1290$ & $-41.7269$ & 2008-02-21 & $550.0$ & G \\
$88$ & Sz133 & Lupus IV & $240.8730$ & $-41.6673$ & 2008-02-24 & $1000.0$ & G \\
$89$ & Sz94 & Lupus III & $241.9570$ & $-39.0746$ & 2008-02-21 & $630.0$ & G \\
$90$ & Sz96 & Lupus III & $242.0520$ & $-39.1425$ & 2008-02-21 & $630.0$ & G \\
$91$ & Sz97 & Lupus III & $242.0910$ & $-39.0726$ & 2008-02-21 & $630.0$ & G \\
 &  &  &  &  & 2008-02-23 & $200.0$ &  \\
 &  &  &  &  & 2008-02-26 & $200.0$ &  \\
$92$ & Sz98 & Lupus III & $242.0940$ & $-39.0794$ & 2008-02-21 & $300.0$ & G \\
 &  &  &  &  & 2008-02-23 & $200.0$ &  \\
 &  &  &  &  & 2008-02-26 & $200.0$ &  \\
$93$ & Sz99 & Lupus III & $242.1000$ & $-39.0970$ & 2008-02-21 & $630.0$ & G \\
 &  &  &  &  & 2008-02-23 & $200.0$ &  \\
 &  &  &  &  & 2008-02-26 & $200.0$ &  \\

%% file: Table_spavslope.tex
$3$ & M6.5 & M4 - M8.5 & $5.30$ & J & $-1.91$ \\
$7$ & M3 & M2 - M4 & $2.50$ & Z & $-1.25$ \\
$13$ & M6.5 & M4 - M8.5 & $2.00$ & J & $-1.54$ \\
$16$ & M5.5 & M4 - M6 & $1.20$ & Z & $-1.21$ & yes\\
$17$ & M5.5 & M4 - M6 & $0.00$ & Z & $  $ \\
$18$ & M5.5 & M4 - M8.5 &   &   &  \\
$19$ & M6 & M4 - M8.5 &   &   &  \\
$20$ & M3.5 & M2 - M4 &   &   &  \\
$23$ & M4.5 & M3 - M6 &   &   &  & yes\\
$24$ & M6 & M4 - M8.5 &   &   &  & yes\\
$25$ & M5.5 & M4 - M6 &   &   &  \\
$26$ & K0 & K0 - K2 &   &   &  \\
$27$ & M6 & M4 - M8.5 & $1.40$ & Z & $-1.01$ \\
$47$ & M5.5 & M4 - M6 &   &   &  \\
$48$ & M5 & M4 - M6 &   &   &  \\
$51$ & M6 & M4 - M8.5 & $1.80$ & J & $-1.08$ & yes\\
$52$ & M4.5 & M3 - M6 &   &   &  & yes\\
$54$ & M3.5 & M2 - M4 &   &   &  & yes\\
$55$ & M8.5 & M6 - M9.5 & $0.00$ & J & $-2.30$ \\
$56$ & M6 & M4 - M8.5 & $1.20$ & J & $-2.36$ \\
$57$ & M7 & M4 - M9 & $2.60$ & J & $-2.27$ \\
$58$ & M5.5 & M3 - M6 & $2.20$ & Z & $-0.17$ \\
$60$ & M9 & M8.5 - M9.5 & $8.00$ & Z & $-2.16$ \\
$61$ & M5.5 & M4 - M6 & $2.00$ & Z & $-2.39$ \\
$62$ & M3.5 & M2 - M4 & $0.60$ & J & $-1.10$ & yes\\
$63$ & M7.5 & M6 - M9 & $3.40$ & Z & $-2.49$ \\
$64$ & M6.5 & M6 - M8.5 & $2.80$ & J & $-1.82$ \\
$65$ & M4 & M3 - M6 & $0.70$ & Z & $-0.21$ \\
$66$ & M6 & M4 - M8.5 & $0.40$ & J & $-1.18$ & yes\\
$67$ & M5.5 & M3 - M6 & $0.00$ & J & $-0.93$ & yes\\
$68$ & M6.5 & M6 - M8.5 & $2.90$ & J & $-2.31$ \\
$69$ & M2.5 & M0.5 - M4 & $1.90$ & J & $-2.30$ \\
$70$ & M6.5 & M4 - M8.5 & $0.00$ & J & $-1.99$ & yes\\
$71$ & K2 & K0 - K7 & $1.60$ & J & $-0.60$ \\
$74$ & M8 & M6 - M8.5 & $1.60$ & J & $-2.34$ \\
$75$ & M8.5 & M6 - M9.5 & $1.00$ & J & $-2.15$ \\
$76$ & M4.5 & M3 - M6 & $1.20$ & Z & $-0.80$ & yes\\
$77$ & M4.5 & M3 - M6 & $0.70$ & Z & $-1.68$ & yes\\
$78$ & K2 & K0 - K7 & $2.50$ & J & $0.59$ \\
$79$ & M4.5 & M3 - M6 & $1.00$ & Z & $-0.86$ & yes\\
$80$ & M5.5 & M3 - M6 & $0.70$ & Z & $-0.91$ & yes\\
$81$ & M6.5 & M4 - M8.5 & $0.00$ & Z & $-1.64$ \\
$82$ & M5.5 & M3 - M6 & $0.80$ & Z & $  $ & yes\\
$83$ & M3 & M0.5 - M4 & $0.20$ & J & $-0.71$ & yes\\
$84$ & M1.5 & K7 - M2 & $2.40$ & Z & $-0.67$ \\
$85$ & M4 & M3 - M6 & $1.30$ & Z & $-0.62$ & yes\\
$86$ & K7 & K2 - M0 & $2.60$ & J & $-0.90$ & yes\\
$87$ & M2 & M0.5 - M4 & $0.60$ & J & $-0.94$ & yes\\
$88$ & K2 & K0 - K7 &   &   &  & yes\\
$89$ & M4.5 & M3 - M6 &   &   &  \\
$90$ & M2 & M0 - M3 & $1.43$ & J & $-0.91$ & yes\\
$91$ & M4.5 & M3 - M6 & $0.30$ & Z & $-1.24$ & yes\\
$92$ & K5 & K2 - M0.5 & $2.50$ & J & $-0.64$ & yes\\
$93$ & M4 & M3 - M6 & $0.40$ & J & $-1.12$ & yes\\

%% file: Table_prop.tex
$3$ & $2935.0^{+335.0}_{-380.0}$ & $2.09^{+0.80}_{-0.61}$ & $0.610^{+0.45}
_{-0.31}$ & $   $ & $   $ & $   $ & $   $ \\[1.5ex]
$7$ & $3415.0^{+145.0}_{-145.0}$ & $0.06^{+0.03}_{-0.02}$ & $0.010^{+0.01}
_{-0.01}$ & $13.64^{+14.82}_{-7.21}$ & $0.35^{+0.15}_{-0.12}$ & $9.90^{+6.37}
_{-3.46}$ & $0.27^{+0.07}_{-0.06}$ \\[1.5ex]
$13$ & $2935.0^{+335.0}_{-380.0}$ & $2.74^{+1.28}_{-0.94}$ & $1.930^{+1.05}
_{-0.74}$ & $   $ & $   $ & $   $ & $   $ \\[1.5ex]
$16$ & $3057.5^{+212.5}_{-67.5}$ & $0.09^{+0.05}_{-0.03}$ & $0.020^{+0.02}
_{-0.02}$ & $1.17^{+2.79}_{-0.78}$ & $0.12^{+0.13}_{-0.02}$ & $3.85^{+0.96}
_{-3.85}$ & $0.14^{+0.08}_{-0.02}$ \\[1.5ex]
$17$ & $3057.5^{+212.5}_{-67.5}$ & $0.01^{+0.00}_{-0.00}$ & $0.000^{+0.01}
_{-0.00}$ & $12.75^{+42.27}_{-4.13}$ & $0.09^{+0.09}_{-0.02}$ & $16.25^{+48.83}
_{-7.41}$ & $0.09^{+0.09}_{-0.02}$ \\[1.5ex]
$27$ & $2990.0^{+280.0}_{-435.0}$ & $0.04^{+0.02}_{-0.02}$ & $0.010^{+0.01}
_{-0.01}$ & $1.87^{+7.16}_{-1.87}$ & $0.09^{+0.13}_{-0.09}$ & $4.43^{+4.99}
_{-2.82}$ & $0.09^{+0.10}_{-0.09}$ \\[1.5ex]
$51$ & $2990.0^{+280.0}_{-435.0}$ & $0.12^{+0.04}_{-0.03}$ & $0.010^{+0.02}
_{-0.01}$ & $0.19^{+2.79}_{-0.19}$ & $0.11^{+0.15}_{-0.11}$ & $3.62^{+0.59}
_{-0.59}$ & $0.17^{+0.06}_{-0.06}$ \\[1.5ex]
$55$ & $2555.0^{+435.0}_{-255.0}$ & $0.17^{+0.08}_{-0.06}$ & $0.050^{+0.05}
_{-0.02}$ & $   $ & $   $ & $   $ & $   $ \\[1.5ex]
$56$ & $2990.0^{+280.0}_{-435.0}$ & $0.37^{+0.17}_{-0.13}$ & $0.060^{+0.08}
_{-0.04}$ & $0.00^{+0.57}_{-0.00}$ & $0.18^{+0.12}_{-0.18}$ & $1.81^{+-1.81}
_{-1.81}$ & $0.26^{+-0.26}_{-0.26}$ \\[1.5ex]
$57$ & $2880.0^{+390.0}_{-480.0}$ & $0.15^{+0.07}_{-0.05}$ & $0.030^{+0.03}
_{-0.02}$ & $0.00^{+2.44}_{-0.00}$ & $0.06^{+5.02}_{-0.00}$ & $3.40^{+-3.40}
_{-3.40}$ & $0.24^{+-0.24}_{-0.24}$ \\[1.5ex]
$58$ & $3057.5^{+357.5}_{-67.5}$ & $0.01^{+0.01}_{-0.00}$ & $0.020^{+0.01}
_{-0.01}$ & $9.83^{+131.97}_{-2.07}$ & $0.09^{+0.20}_{-0.02}$ & $13.83^{+101.47}
_{-5.94}$ & $0.09^{+0.16}_{-0.02}$ \\[1.5ex]
$60$ & $2400.0^{+155.0}_{-100.0}$ & $1.94^{+0.95}_{-0.66}$ & $0.800^{+0.61}
_{-0.40}$ & $   $ & $   $ & $   $ & $   $ \\[1.5ex]
$61$ & $3057.5^{+212.5}_{-67.5}$ & $1.22^{+0.60}_{-0.42}$ & $0.190^{+0.33}
_{-0.19}$ & $   $ & $   $ & $   $ & $   $ \\[1.5ex]
$62$ & $3342.5^{+217.5}_{-72.5}$ & $0.16^{+0.06}_{-0.05}$ & $0.060^{+0.03}
_{-0.03}$ & $2.86^{+5.43}_{-0.69}$ & $0.31^{+0.20}_{-0.05}$ & $3.49^{+1.67}
_{-0.67}$ & $0.27^{+0.09}_{-0.03}$ \\[1.5ex]
$63$ & $2795.0^{+195.0}_{-395.0}$ & $2.93^{+1.43}_{-1.00}$ & $1.140^{+0.94}
_{-0.61}$ & $   $ & $   $ & $   $ & $   $ \\[1.5ex]
$64$ & $2935.0^{+55.0}_{-380.0}$ & $1.56^{+0.60}_{-0.45}$ & $0.450^{+0.34}
_{-0.23}$ & $   $ & $   $ & $   $ & $   $ \\[1.5ex]
$65$ & $3270.0^{+145.0}_{-280.0}$ & $0.00^{+0.00}_{-0.00}$ & $0.000^{+0.01}
_{-0.00}$ & $   $ & $   $ & $   $ & $   $ \\[1.5ex]
$66$ & $2990.0^{+280.0}_{-435.0}$ & $0.03^{+0.01}_{-0.01}$ & $0.000^{+0.01}
_{-0.00}$ & $3.07^{+12.27}_{-3.07}$ & $0.09^{+0.12}_{-0.09}$ & $5.02^{+9.65}
_{-2.91}$ & $0.08^{+0.10}_{-0.08}$ \\[1.5ex]
$67$ & $3057.5^{+357.5}_{-67.5}$ & $0.09^{+0.04}_{-0.03}$ & $0.020^{+0.02}
_{-0.01}$ & $1.17^{+7.21}_{-0.78}$ & $0.12^{+0.24}_{-0.02}$ & $3.85^{+2.80}
_{-3.85}$ & $0.14^{+0.14}_{-0.02}$ \\[1.5ex]
$68$ & $2935.0^{+55.0}_{-380.0}$ & $7.12^{+2.72}_{-2.07}$ & $0.990^{+1.27}
_{-0.82}$ & $   $ & $   $ & $   $ & $   $ \\[1.5ex]
$69$ & $3487.5^{+290.0}_{-217.5}$ & $0.13^{+0.05}_{-0.04}$ & $0.020^{+0.02}
_{-0.02}$ & $7.98^{+19.47}_{-5.19}$ & $0.44^{+0.26}_{-0.18}$ & $5.74^{+8.90}
_{-1.79}$ & $0.33^{+0.19}_{-0.09}$ \\[1.5ex]
$70$ & $2935.0^{+335.0}_{-380.0}$ & $0.06^{+0.02}_{-0.02}$ & $0.000^{+0.01}
_{-0.00}$ & $0.44^{+7.07}_{-0.44}$ & $0.08^{+0.15}_{-0.08}$ & $3.57^{+4.02}
_{-2.32}$ & $0.08^{+0.12}_{-0.08}$ \\[1.5ex]
$71$ & $4900.0^{+350.0}_{-840.0}$ & $0.04^{+0.02}_{-0.01}$ & $0.020^{+0.01}
_{-0.01}$ & $   $ & $   $ & $   $ & $   $ \\[1.5ex]
$74$ & $2710.0^{+280.0}_{-155.0}$ & $2.44^{+0.93}_{-0.71}$ & $0.380^{+0.43}
_{-0.29}$ & $   $ & $   $ & $   $ & $   $ \\[1.5ex]
$75$ & $2555.0^{+435.0}_{-255.0}$ & $2.30^{+0.88}_{-0.67}$ & $0.570^{+0.47}
_{-0.31}$ & $   $ & $   $ & $   $ & $   $ \\[1.5ex]
$76$ & $3197.5^{+217.5}_{-207.5}$ & $0.27^{+0.13}_{-0.09}$ & $0.100^{+0.09}
_{-0.05}$ & $0.53^{+1.73}_{-0.53}$ & $0.20^{+0.20}_{-0.04}$ & $1.94^{+0.75}
_{-1.94}$ & $0.23^{+0.08}_{-0.02}$ \\[1.5ex]
$77$ & $3197.5^{+217.5}_{-207.5}$ & $0.41^{+0.20}_{-0.14}$ & $0.060^{+0.10}
_{-0.06}$ & $0.15^{+1.25}_{-0.15}$ & $0.19^{+0.26}_{-0.01}$ & $0.82^{+1.34}
_{-0.82}$ & $0.25^{+0.07}_{-0.02}$ \\[1.5ex]
$78$ & $4900.0^{+350.0}_{-840.0}$ & $0.03^{+0.01}_{-0.01}$ & $0.130^{+0.04}
_{-0.03}$ & $   $ & $   $ & $   $ & $   $ \\[1.5ex]
$79$ & $3197.5^{+217.5}_{-207.5}$ & $0.20^{+0.10}_{-0.07}$ & $0.060^{+0.06}
_{-0.04}$ & $0.99^{+1.86}_{-0.98}$ & $0.20^{+0.18}_{-0.07}$ & $2.48^{+0.67}
_{-2.48}$ & $0.22^{+0.09}_{-0.04}$ \\[1.5ex]
$80$ & $3057.5^{+357.5}_{-67.5}$ & $0.13^{+0.06}_{-0.04}$ & $0.040^{+0.04}
_{-0.02}$ & $0.64^{+5.03}_{-0.51}$ & $0.13^{+0.24}_{-0.01}$ & $2.82^{+1.96}
_{-2.82}$ & $0.15^{+0.14}_{-0.02}$ \\[1.5ex]
$81$ & $2935.0^{+335.0}_{-380.0}$ & $0.15^{+0.07}_{-0.05}$ & $0.010^{+0.04}
_{-0.01}$ & $0.00^{+2.36}_{-0.00}$ & $0.10^{+0.16}_{-0.10}$ & $3.28^{+-3.28}
_{-3.28}$ & $0.24^{+-0.24}_{-0.24}$ \\[1.5ex]
$82$ & $3057.5^{+357.5}_{-67.5}$ & $0.09^{+0.05}_{-0.03}$ & $0.030^{+0.03}
_{-0.02}$ & $1.18^{+7.23}_{-0.79}$ & $0.12^{+0.24}_{-0.02}$ & $3.86^{+2.82}
_{-3.86}$ & $0.14^{+0.14}_{-0.02}$ \\[1.5ex]
$83$ & $3415.0^{+362.5}_{-145.0}$ & $0.26^{+0.10}_{-0.08}$ & $0.090^{+0.06}
_{-0.04}$ & $2.35^{+7.10}_{-1.20}$ & $0.40^{+0.38}_{-0.12}$ & $2.73^{+2.25}
_{-0.38}$ & $0.31^{+0.20}_{-0.06}$ \\[1.5ex]
$84$ & $3632.5^{+427.5}_{-72.5}$ & $0.11^{+0.05}_{-0.04}$ & $0.040^{+0.03}
_{-0.03}$ & $19.64^{+59.62}_{-9.80}$ & $0.58^{+0.08}_{-0.08}$ & $9.95^{+50.10}
_{-3.95}$ & $0.40^{+0.24}_{-0.04}$ \\[1.5ex]
$85$ & $3270.0^{+145.0}_{-280.0}$ & $0.49^{+0.24}_{-0.17}$ & $0.220^{+0.16}
_{-0.11}$ & $0.26^{+0.81}_{-0.26}$ & $0.32^{+0.16}_{-0.10}$ & $1.12^{+0.96}
_{-1.12}$ & $0.27^{+0.05}_{-0.01}$ \\[1.5ex]
$86$ & $4060.0^{+840.0}_{-210.0}$ & $0.92^{+0.35}_{-0.27}$ & $0.560^{+0.27}
_{-0.21}$ & $3.72^{+20.71}_{-1.37}$ & $1.16^{+0.04}_{-0.16}$ & $2.54^{+13.53}
_{-1.09}$ & $0.76^{+0.39}_{-0.20}$ \\[1.5ex]
$87$ & $3560.0^{+217.5}_{-290.0}$ & $0.21^{+0.10}_{-0.07}$ & $0.060^{+0.06}
_{-0.03}$ & $5.77^{+7.30}_{-4.20}$ & $0.52^{+0.23}_{-0.25}$ & $3.74^{+3.02}
_{-1.03}$ & $0.37^{+0.14}_{-0.12}$ \\[1.5ex]
$90$ & $3560.0^{+290.0}_{-145.0}$ & $0.82^{+0.31}_{-0.24}$ & $0.270^{+0.19}
_{-0.13}$ & $0.88^{+1.79}_{-0.56}$ & $0.67^{+0.32}_{-0.08}$ & $0.81^{+0.94}
_{-0.81}$ & $0.39^{+0.18}_{-0.06}$ \\[1.5ex]
$91$ & $3197.5^{+217.5}_{-207.5}$ & $0.18^{+0.09}_{-0.06}$ & $0.040^{+0.05}
_{-0.03}$ & $1.29^{+1.86}_{-1.27}$ & $0.21^{+0.16}_{-0.08}$ & $2.70^{+1.00}
_{-2.70}$ & $0.22^{+0.09}_{-0.05}$ \\[1.5ex]
$92$ & $4350.0^{+550.0}_{-572.5}$ & $2.35^{+0.90}_{-0.68}$ & $1.180^{+0.66}
_{-0.47}$ & $2.42^{+3.02}_{-2.24}$ & $1.53^{+0.39}_{-0.45}$ & $1.49^{+3.46}
_{-1.49}$ & $1.12^{+0.54}_{-0.65}$ \\[1.5ex]
$93$ & $3270.0^{+145.0}_{-280.0}$ & $0.11^{+0.04}_{-0.03}$ & $0.030^{+0.02}
_{-0.02}$ & $3.07^{+3.72}_{-2.85}$ & $0.26^{+0.11}_{-0.15}$ & $4.34^{+1.22}
_{-4.34}$ & $0.23^{+0.06}_{-0.10}$ \\[1.5ex]

%% file: Table_ha.tex
$7$ & $201.32$ & yes & $351.0\pm5.0$ & yes & $-9.49\pm0.39$ &   \\
$13$ & $4.15$ &   & $98.0\pm11.0$ &   & $-11.94\pm0.33$ &   \\
$16$ & $27.16$ & yes & $306.0\pm4.0$ & yes & $-9.92\pm0.37$ & yes \\
$17$ & $4.66$ &   & $94.0\pm11.5$ &   & $-11.98\pm0.33$ &   \\
$18$ & $7.07$ &   & $116.0\pm7.5$ &   & $-11.76\pm0.32$ &   \\
$19$ & $2.23$ &   & $100.0\pm11.0$ &   & $-11.92\pm0.33$ & yes \\
$20$ & $8.24$ &   & $137.0\pm16.5$ &   & $-11.56\pm0.35$ &   \\
$22$ & $327.34$ & yes & $312.0\pm4.5$ & yes & $-9.86\pm0.37$ & yes \\
$23$ & $5.20$ &   & $183.0\pm16.0$ &   & $-11.11\pm0.36$ & yes \\
$24$ & $10.62$ &   & $115.0\pm6.0$ &   & $-11.77\pm0.32$ &   \\
$25$ & $7.01$ &   & $112.0\pm7.0$ &   & $-11.80\pm0.32$ &   \\
$27$ & $36.56$ &   & $137.0\pm5.5$ &   & $-11.56\pm0.32$ &   \\
$47$ & $9.77$ &   & $104.0\pm8.0$ &   & $-11.88\pm0.32$ & yes \\
$48$ & $3.37$ &   & $159.0\pm7.0$ &   & $-11.35\pm0.33$ &   \\
$49$ & $44.74$ & yes & $224.0\pm19.5$ & yes & $-10.72\pm0.39$ & yes \\
$50$ & $147.24$ & yes & $385.0\pm6.5$ & yes & $-9.16\pm0.41$ & yes \\
$51$ & $11.55$ &   & $125.0\pm6.0$ &   & $-11.68\pm0.32$ &   \\
$52$ & $130.76$ & yes & $289.0\pm4.0$ & yes & $-10.09\pm0.36$ &   \\
$54$ & $26.84$ & yes & $501.0\pm6.0$ & yes & $-8.03\pm0.47$ & yes \\
$58$ & $10.66$ &   & $139.0\pm11.0$ &   & $-11.54\pm0.33$ & yes \\
$59$ & $255.62$ & yes & $264.0\pm8.0$ & yes & $-10.33\pm0.36$ &   \\
$62$ & $22.29$ & yes & $431.0\pm10.5$ & yes & $-8.71\pm0.44$ & yes \\
$65$ & $17.68$ &   & $463.0\pm27.5$ & yes & $-8.40\pm0.52$ & yes \\
$66$ & $23.59$ &   & $181.0\pm6.0$ &   & $-11.13\pm0.33$ &   \\
$67$ & $9.43$ &   & $149.0\pm6.0$ &   & $-11.44\pm0.32$ &   \\
$70$ & $11.93$ &   & $123.0\pm6.5$ &   & $-11.70\pm0.32$ &   \\
$71$ & $0.68$ &   & $167.0\pm24.0$ &   & $-11.27\pm0.40$ &   \\
$76$ & $97.79$ & yes & $348.0\pm4.5$ & yes & $-9.51\pm0.39$ &   \\
$77$ & $23.74$ & yes & $504.0\pm9.0$ & yes & $-8.00\pm0.47$ &   \\
$78$ & $333.58$ & yes & $283.0\pm7.0$ & yes & $-10.14\pm0.37$ &   \\
$79$ & $6.37$ &   & $171.0\pm9.5$ &   & $-11.23\pm0.34$ &   \\
$80$ & $26.84$ & yes & $182.0\pm4.5$ &   & $-11.12\pm0.33$ &   \\
$81$ & $10.54$ &   & $236.0\pm6.0$ & yes & $-10.60\pm0.35$ &   \\
$82$ & $68.76$ & yes & $389.0\pm6.0$ & yes & $-9.12\pm0.41$ & yes \\
$83$ & $71.81$ & yes & $440.0\pm5.5$ & yes & $-8.62\pm0.43$ &   \\
$84$ & $267.22$ & yes & $450.0\pm6.5$ & yes & $-8.52\pm0.44$ & yes \\
$85$ & $139.22$ & yes & $261.0\pm3.5$ & yes & $-10.36\pm0.35$ &   \\
$86$ & $40.99$ & yes & $486.0\pm6.0$ & yes & $-8.18\pm0.46$ & yes \\
$87$ & $58.32$ & yes & $229.0\pm5.0$ & yes & $-10.67\pm0.34$ &   \\
$88$ & $121.73$ & yes & $498.0\pm4.0$ & yes & $-8.06\pm0.46$ & yes \\
$89$ & $7.49$ &   & $156.0\pm6.0$ &   & $-11.38\pm0.32$ &   \\
$90$ & $11.40$ & yes & $344.0\pm6.5$ & yes & $-9.55\pm0.39$ &   \\
$91$ & $38.67$ & yes & $395.0\pm6.5$ & yes & $-9.06\pm0.41$ &   \\
$92$ & $26.42$ & yes & $432.0\pm4.0$ & yes & $-8.70\pm0.43$ & yes \\
$93$ & $116.96$ & yes & $379.0\pm3.0$ & yes & $-9.21\pm0.40$ &   \\

%% file: FinalDraft.bbl
\begin{thebibliography}{99}

\bibitem[Alcal{\'a} et al.(2011)]{Alcala11} Alcal{\'a}, J.~M., 
et al.\ 2011, Astronomische Nachrichten, 332, 242 
\bibitem[Allard et al.(2000)]{Allard00} Allard, F., Hauschildt,
  P.~H., Alexander, D.~R., Ferguson, J.~W., \& Tamanai, A.\ 2000, From
  Giant Planets to Cool Stars, 212, 127
\bibitem[Allen et al.(2007)]{Allen07} Allen, P.~R., Luhman, 
K.~L., Myers, P.~C., Megeath, S.~T., Allen, L.~E., Hartmann, L., 
\& Fazio, G.~G.\ 2007, \apj, 655, 1095
\bibitem[Allers et al.(2006)]{Allers06} Allers, K.~N., Kessler-Silacci, 
J.~E., Cieza, L.~A., \& Jaffe, D.~T.\ 2006, \apj, 644, 364
\bibitem[Baraffe et al.(1998)]{Baraffe98} Baraffe, I., Chabrier, G.,
  Allard, F., \& Hauschildt, P.~H.\ 1998, \aap, 337, 403
\bibitem[Baraffe et al.(2001)]{Baraffe01} Baraffe, I., Chabrier, G.,
  Allard, F., \& Hauschildt, P.\ 2001, From Darkness to Light: Origin
  and Evolution of Young Stellar Clusters, 243, 571
\bibitem[Barrado y Navascu{\'e}s \& Mart{\'{\i}}n(2003)]{Bar03} 
Barrado y Navascu{\'e}s, D., \& Mart{\'{\i}}n, E.~L.\ 2003, \aj, 126, 2997
\bibitem[Comer{\'o}n(2008)]{Comeron08} Comer{\'o}n, F.\ 2008,
  Handbook of Star Forming Regions, Volume II, 295
\bibitem[Comer{\'o}n et al.(2009)]{Comeron09} Comer{\'o}n, F.,
  Spezzi, L., \& L{\'o}pez Mart{\'{\i}}, B.\ 2009, \aap, 500, 1045
\bibitem[Evans et al.(2003)]{Evans03} Evans, N.~J., II, et al.\ 2003,
  \pasp, 115, 965
\bibitem[Evans et al.(2007)]{Evans07} Evans, N.~J., et al.\ 2007,
  Final Delivery of Data from the c2d Legacy Project: IRAC and MIPS
  (Pasadena: SSC)\footnote{http://ssc.spitzer.caltech.edu/spitzermission/observingprograms/legacy/c2dhistory.html}
\bibitem[Evans et al.(2009)]{Evans09} Evans, N.~J., et al.\ 2009,
  \apjs, 181, 321
\bibitem[Harvey et al.(2007a)]{Harvey07a} Harvey, P.~M., et al.\ 2007,
  \apj, 663, 1139
\bibitem[Harvey et al.(2007b)]{Harvey07b} Harvey, P., Mer{\'{\i}}n,
  B., Huard, T.~L., Rebull, L.~M., Chapman, N., Evans, N.~J., II, \&
  Myers, P.~C.\ 2007, \apj, 663, 1149
\bibitem[Hauschildt et al.(1999)]{Haus99} Hauschildt, P.~H., Allard,
  F., Ferguson, J., Baron, E., \& Alexander, D.~R.\ 1999, \apj, 525,
  871
\bibitem[Hu{\'e}lamo et al.(2010)]{Hue10} Hu{\'e}lamo, N., 
et al.\ 2010, \aap, 523, A42
\bibitem[Hughes et al.(1994)]{Hughes94} Hughes, J., Hartigan, 
P., Krautter, J., \& Kelemen, J.\ 1994, \aj, 108, 1071
\bibitem[Jayawardhana et al.(2003)]{Jaya03} Jayawardhana, R.,
  Mohanty, S., \& Basri, G.\ 2003, \apj, 592, 282
\bibitem[Kenyon \& Hartmann(1995)]{Kenyon95} Kenyon, S.~J., \&
  Hartmann, L.\ 1995, \apjs, 101, 117
\bibitem[L{\'o}pez Mart{\'{\i}} et al.(2005)]{Lopez05} L{\'o}pez 
Mart{\'{\i}}, B., Eisl{\"o}ffel, J., \& Mundt, R.\ 2005, \aap, 440, 139
\bibitem[Luhman et al.(2003)]{Luhman03} Luhman, K.~L., Stauffer,
  J.~R., Muench, A.~A., Rieke, G.~H., Lada, E.~A., Bouvier, J., \&
  Lada, C.~J.\ 2003, \apj, 593, 1093
\bibitem[Luhman(2007)]{Luhman07} Luhman, K.~L.\ 2007, \apjs, 
173, 104
\bibitem[Mer{\'{\i}}n et al.(2007)]{Merin07} Mer{\'{\i}}n, B., 
et al.\ 2007, \apj, 661, 361
\bibitem[Mer{\'\i}n et al.(2008)]{Merin08} Mer{\'{\i}}n, B., et
  al.\ 2008, \apjs, 177, 551
\bibitem[Mer{\'{\i}}n et al.(2010)]{Merin10} Mer{\'{\i}}n, B., et
  al.\ 2010, \apj, 718, 1200
\bibitem[Montes(1998)]{Montes98} Montes, D.\ 1998, \apss, 263, 275
\bibitem[Natta et al.(2004)]{Natta04} Natta, A., Testi, L.,
  Muzerolle, J., Randich, S., Comer{\'o}n, F., \& Persi, P.\ 2004,
  \aap, 424, 603
\bibitem[Natta et al.(2005)]{Natta05} Natta, A., Testi, L., 
Randich, S., \& Muzerolle, J.\ 2005, \memsai, 76, 343 
\bibitem[Oliveira et al.(2009)]{Oliveira09} Oliveira, I., et
  al.\ 2009, \apj, 691, 672
\bibitem[Sicilia-Aguilar et al.(2010)]{Sicilia10} 
Sicilia-Aguilar, A., Henning, T., \& Hartmann, L.~W.\ 2010, \apj, 710, 597 
\bibitem[Siess et al.(2000)]{Siess00} Siess, L., Dufour, E., \&
  Forestini, M.\ 2000, \aap, 358, 593
\bibitem[Tothill et al.(2009)]{Tothill09} Tothill, N.~F.~H., et 
al.\ 2009, \apjs, 185, 98
\bibitem[Wang \& Henning(2009)]{WH09} Wang, H., \& Henning, T.\ 2009,
  \aj, 138, 1072
\bibitem[Weingartner \& Draine(2001)]{WD01} Weingartner, J.~C., \&
  Draine, B.~T.\ 2001, \apj, 548, 296
\bibitem[White \& Basri(2003)]{WB03} White, R.~J., \& Basri,
  G.\ 2003, \apj, 582, 1109

\end{thebibliography}
